\documentclass[nofootinbib,superscriptaddress,onecolumn,amsmath,amssymb,showpacs,showkeys,notitlepage,aps,prd]{revtex4-1}
\usepackage[utf8]{inputenc} 
\usepackage{subfig}
\usepackage[T1]{fontenc}

\usepackage{mathrsfs}  
\usepackage{cases}
\usepackage{physics}
\usepackage{bm}
\usepackage{academicons}
\usepackage{mathtools, nccmath}
\usepackage{fancyhdr}
\usepackage{tikz,xcolor}

\usepackage{tensor}
\usepackage[normalem]{ulem}
\usepackage{lipsum}
\usepackage{soul}
\usepackage{cancel}
\usepackage{stackengine,scalerel}
\usepackage{hyperref}
\usepackage{tabularx}
\usepackage[justification=raggedright]{caption}
\usepackage{placeins}
\hypersetup{colorlinks, linkcolor={red},citecolor={blue},urlcolor={blue}}  

\usepackage{orcidlink}
\usepackage{graphicx}

\usepackage[utf8]{inputenc}
\usepackage{amsmath,amssymb,amsfonts}
\usepackage{psfrag}
\usepackage{makeidx}
\usepackage{bm}
\usepackage{epsf}
\usepackage{float}
\usepackage{multirow}
\usepackage{booktabs}
\usepackage{array}
\usepackage{slashed}

\newcommand{\rH}{r_H}
\newcommand{\rtopo}{r_{topo}}

\newcommand{\Mcrit}{M_{\rm crit}}

\newcommand{\Dtilde}{\tilde\nabla}

\begin{document}
\pagestyle{plain}

\title{Exotic Spinor Deformation of a Schwarzschild Black Hole: Perturbations and Stability}
\author{Lu\'{i}s Rodolfo dos Santos Filho}
\email[]{luis.um.dia.seja@gmail.com}
\affiliation{Faculty of Engineering and Science of Guaratinguet\'{a}, UNESP, Mathematics Department, S\~{a}o Paulo, Brazil}
\affiliation{S\~{a}o Paulo State Department of Education (SEDUC-SP), S\~{a}o Paulo, Brazil}

\author{Bertha Cuadros-Melgar}
\email[]{bertha@usp.br}
\affiliation{Engineering School of Lorena - University of Sao Paulo (EEL-USP), Estrada Municipal do Campinho Nº 100, Campinho, CEP: 12602-810, Lorena, SP - Brazil}

\begin{abstract}

 In this work we investigate how exotic spinorial topology modifies the propagation of fermionic modes in the Schwarzschild geometry. By incorporating a topological gradient, $\partial_\mu\varphi$, into the Dirac operator through a tetrad deformation, we show that the exotic spinorial sector naturally induces an effective geometric deformation of the Schwarzschild background. The central consequence of this construction is the emergence of a topological impedance surface at $\rtopo=1/q$, which suppresses the radial group velocity of the exotic spinorial modes, yielding a birefringent propagation pattern. Using a spinorial perturbation we discuss the greybody factor and the quasinormal response. The quasinormal spectrum shows the stability of the background metric and the absence of isospectrality for the superpartner potentials.

\end{abstract}

{\let\clearpage\relax \maketitle}

\section{Introduction}
\label{sec:intro}

The description of spinor fields in curved spacetime has traditionally been explored through two complementary paths: the algebraic approach, rooted in the Clifford structure of Dirac matrices, and the geometric perspective championed by Cartan, where spacetime points emerge from spinorial condensates \cite{cartan1981theory}. Alongside these fundamental discussions, the existence of exotic spinor structures on manifolds that are not simply connected, $H^{1}(M,\mathbb Z_2)\neq0$, has opened a window into how global topological data can manifest as local dynamical corrections \cite{Balachandran:1993ts,HoffdaSilva:2009aa,HoffdaSilva2016}.

In recent years, the geometrization of topology program has shown that topological information can be incorporated into the Dirac operator through a tetrad-like formalism \cite{HoffdaSilva:2022xx}. By redefining the Dirac matrices to include a topological gradient, $\partial_\mu\varphi$, the modified Clifford structure naturally defines an effective metric $\tilde g^{\mu\nu}$. In flat spacetime regimes, this construction gives rise to spin-orbit effects analogous to Rashba-type interactions and to dissipative behavior associated with quasinormal-mode dynamics, without requiring external gauge fields \cite{dosSantosFilho:2026epjc}.

In the present work, we promote this effective metric to a semiclassical spacetime geometry induced by the exotic spinorial/topological sector. Therefore, the deformation is not interpreted as a metric privately perceived by exotic spinors. Rather, the parameter $q=\partial_r \varphi$ is treated as an effective spinorial/topological hair whose stress-energy content modifies the radial geometry.
Despite these advances, the consequences of this geometrization remain largely unexplored in strong gravitational fields. The Schwarzschild black hole provides a clean testbed, since its causal structure is simple and its thermodynamics is well understood. In such a background, the modified Clifford algebra is expected to alter the radial propagation of exotic spinorial excitations and, consequently, the effective potential barriers, greybody factors, and quasinormal response of spinorial perturbations.

Quasinormal modes (QNMs) are defined as damped oscillations that are independent of the initial perturbation. Characterized solely by the parameters of the background metric, their complex frequencies can be regarded as the unique footprints of a given spacetime geometry~\cite{Nollert:1999ji,Kokkotas:1999bd,Konoplya:2011qq}. QNMs corresponding to various wave equations have been extensively studied since the seminal work of Regge and Wheeler~\cite{Regge:1957td}, which pioneered the stability analysis of the Schwarzschild black hole under metric perturbations. Although most studies focus on scalar perturbations due to their mathematical simplicity and cosmological applications, investigating other fields—such as spinor fields—is particularly appealing~\cite{Cho:2003qe,Zhidenko:2003wq}. Their richer structure and additional degrees of freedom can unveil novel physical phenomena; for instance, unlike boson fields, fermion fields do not undergo superradiant scattering in Kerr or Reissner--Nordstr\"om spacetimes~\cite{Unruh:1973ms,Chandrasekhar:1976zp,Guven:1977,Soffel:1977,Konoplya:2007_KNdS}. Dirac perturbations have also been studied in higher dimensional models~\cite{Cho:2007,LopezOrtega:2009}, $(2+1)$--gravity~\cite{PhysRevD.63.124015,CuadrosMelgar:2012,CuadrosMelgar:2022}, string-inspired black holes~\cite{FernandezPiedra:2011}, and braneworld theories~\cite{Gibbons:2008_brane,Cho:2008}. 

In the framework of modified gravity theories, spinorial perturbations are especially useful probes and their emission spectra can reveal distinct signatures of non-standard fermionic fields \cite{Cavalcanti:2015raa,Gibbons:2008,Li:2013,Fernando:2010}. In the context of Galileon black holes, for example, vectorial and spinorial perturbations have been used to diagnose quasinormal spectra, quasiresonant regimes, and linear stability \cite{Abdalla:2018cmx}. This provides a useful precedent for the present analysis: here, the role played by the Galileon coupling is replaced by the spinorial/topological hair parameter \(q\), and the near-critical behavior of the damping rate is used as a diagnostic of whether the system approaches a long-lived quasiresonant regime or develops an instability.

The central aim of this work is to investigate how exotic spinorial topology can deform the thermodynamic properties and perturbative behavior of Schwarzschild black holes. We propose that a radial spinorial/topological hair modifies the effective radial geometry and acts as a birefringent filter for spinorial perturbations near the horizon. This effect does not require the introduction of an additional Killing horizon. Instead, it arises from an impedance surface in the radial sector, which suppresses the propagation and transmission of topologically deformed spinorial modes. Since the semiclassical geometry has a modified radial coefficient, the corresponding surface gravity and effective temperature are shifted relative to the Schwarzschild value.

This paper is organized as follows. Section~\ref{sec:metric} derives the effective topological deformation of the Schwarzschild geometry, reconstructs the anisotropic source supporting it, and characterizes the topological impedance surface. In Section~\ref{sec:friction} we analyze the resulting birefringent dispersion relation and the suppression of radial group velocity. Section~\ref{sec:dirac_perturbations} is devoted to the black hole response to linear exotic Dirac perturbations through the quasinormal modes produced by topologically deformed superpartner potentials. In Section~\ref{sec:numerical_diagnostics} we present the numerical diagnostics: the potential barriers and the corresponding quasinormal frequencies. Section~\ref{sec:hawking} interprets the Hawking emission as a greybody-filter effect, and Section~\ref{sec:remnants} discusses the possible formation of long-lived topological remnant candidates. Lastly, we discuss our final remarks in Section~\ref{sec:concl}. 

\section{Effective Schwarzschild Geometry and Topological Impedance}
\label{sec:metric}

To explore the thermodynamic implications of exotic spinorial topology, it is necessary to specify how the topological sector deforms the underlying spacetime geometry. We start from the Schwarzschild line element,

\begin{equation}
    ds^2=-f(r)dt^2+\frac{dr^2}{f(r)}+r^2d\Omega^2,
    \qquad
    f(r)=1-\frac{2M}{r},
    \label{eq:schwarz_metric}
\end{equation}
where the gravitational event horizon is located at $\rH=2M$. Natural units $G=c=\hbar=k_B=1$ are used throughout.

\subsection{Effective metric from the exotic tetrad}

In the geometrization of topology program, the topological correction is implemented directly in the local frame. In a local inertial patch, the tetrad deformation may be written as,
\begin{equation}
    \tilde e^{a}_{\ \mu}
    =
    \delta^{a}_{\ \mu}-x_\mu\partial^a\varphi,
    \label{eq:tetrad_flat}
\end{equation}
where $\varphi$ is the topological field encoding the non-trivial spinorial structure \cite{dosSantosFilho:2026epjc}. The use of $\varphi$ avoids confusion with the angular Schwarzschild coordinate $\theta$, a symbol employed in the previous reference. The effective metric is obtained by contracting the deformed tetrads with the tangent-space metric as follows,
\begin{align}
    \tilde g_{\mu\nu}
    &=
    \eta_{ab}\tilde e^a_{\ \mu}\tilde e^b_{\ \nu} \nonumber\\
    &=
    \eta_{\mu\nu}
    -x_\mu\partial_\nu\varphi
    -x_\nu\partial_\mu\varphi
    +x_\mu x_\nu(\partial\varphi)^2 \nonumber\\
    &=
    \eta_{\mu\nu}
    -2x_{(\mu}\partial_{\nu)}\varphi
    +x_\mu x_\nu(\partial\varphi)^2.
    \label{eq:metric_flat}
\end{align}
Equation~\eqref{eq:metric_flat} shows that the exotic spinorial coupling naturally defines a disformal geometric structure associated with the modified Clifford algebra.

To extend this construction to the Schwarzschild background, we use the local-inertial-frame interpretation of Eq.~\eqref{eq:metric_flat}. At each spacetime point the topological deformation is defined in the tangent frame and then mapped back to the curved coordinates. At the level of the effective field theory used here, this motivates a semiclassical deformation of the Schwarzschild geometry controlled by the radial spinorial/topological hair. The corresponding inverse effective metric is given by,
\begin{equation}
    \tilde g^{\mu\nu}
    =
    g^{\mu\nu}
    -\left(x^\mu\partial^\nu\varphi+x^\nu\partial^\mu\varphi\right)
    +x^\mu x^\nu(\partial\varphi)^2.
    \label{eq:eff_metric_tensor}
\end{equation}
This object is equivalently defined through the modified Clifford algebra,
\begin{equation}
    \{\tilde\gamma^\mu,\tilde\gamma^\nu\}
    =
    2\tilde g^{\mu\nu}\mathbb I,
    \label{eq:mod_clifford}
\end{equation}
where $\tilde\gamma^\mu=\tilde e^\mu_{\ a}\gamma^a$. Thus, the effective metric is first suggested by the algebra of the exotic Dirac operator. In the gravitational interpretation adopted below, however, it is promoted to a semiclassical spacetime geometry supported by an effective spinorial/topological source.

\subsection{Radial topological gradient and impedance surface}

Spherical symmetry motivates the choice of a purely radial topological gradient,
\begin{equation}
    \partial_\mu\varphi=(0,q,0,0),
    \qquad
    q=\partial_r\varphi,
    \label{eq:radial_gradient}
\end{equation}
with $q$ treated as approximately constant in the radial region of interest. Raising the index with the Schwarzschild background gives,
\begin{equation}
    \partial^r\varphi=g^{rr}\partial_r\varphi=f(r)q,
    \qquad
    \partial_\mu\varphi \partial^\mu\varphi = f(r)q^2
\end{equation}

Taking $x^r=r$ in Eq.~\eqref{eq:eff_metric_tensor} the radial component becomes,
\begin{align}
    \tilde g^{rr}
    &=
    g^{rr}-2r\partial^r\varphi+r^2(\partial\varphi)^2       \nonumber\\
    &=
    f(r)-2rf(r)q+r^2f(r)q^2                                  \nonumber\\
    &=
    f(r)(1-qr)^2.
    \label{eq:grr_eff}
\end{align}
This is the central result of the proposed model.

\subsection{Effective anisotropic source supporting the topological geometry}
\label{subsec:effective_anisotropic_source}

The interpretation of the metric deformation deserves some care. Once the deformation is promoted to a spacetime geometry, it should not be understood as a private metric perceived only by exotic spinors. Rather, it must be associated with an effective source in the gravitational field equations. We therefore reinterpret \(q\) as an effective spinorial/topological hair whose stress-energy content supports the modified radial geometry. A detailed calculation of the relevant geometric quantities is shown in Appendix \ref{app:geometric}.

The effective line element considered in this work is then given by,
\begin{equation}
    d\tilde s^2
    =
    -f(r)dt^2
    +
    \frac{dr^2}{f(r)(1-qr)^2}
    +
    r^2d\Omega^2,
    \qquad
    f(r)=1-\frac{2M}{r}.
    \label{eq:topological_geometry_source}
\end{equation}

The source required to sustain this geometry is obtained from,
\begin{equation}
    G^\mu_{\ \nu}
    =
    8\pi T^\mu_{\ \nu}{}^{\rm topo}.
    \label{eq:einstein_effective_source}
\end{equation}
A direct computation gives the non-vanishing mixed components,
\begin{equation}
    G^t_{\ t}
    =
    \frac{q}{r^2}
    \left[
        -4Mqr+4M+3qr^2-4r
    \right],
    \label{eq:Gtt_effective_source}
\end{equation}
\begin{equation}
    G^r_{\ r}
    =
    \frac{q}{r}(qr-2),
    \label{eq:Grr_effective_source}
\end{equation}
and
\begin{equation}
    G^\theta_{\ \theta}
    =
    G^\phi_{\ \phi}
    =
    \frac{q(r-M)(qr-1)}{r^2}.
    \label{eq:Gang_effective_source}
\end{equation}
Thus, the effective source may be written as an anisotropic fluid,
\begin{equation}
    T^\mu_{\ \nu}{}^{\rm topo}
    =
    \mathrm{diag}
    \left(
        -\rho,
        p_r,
        p_t,
        p_t
    \right),
    \label{eq:anisotropic_source_def}
\end{equation}
where
\begin{equation}
    \rho(r)
    =
    \frac{q}{8\pi r^2}
    \left[
        4r-4M+4Mqr-3qr^2
    \right],
    \label{eq:rho_effective_source}
\end{equation}
\begin{equation}
    p_r(r)
    =
    \frac{q}{8\pi r}
    (qr-2),
    \label{eq:pr_effective_source}
\end{equation}
and
\begin{equation}
    p_t(r)
    =
    \frac{q(r-M)(qr-1)}{8\pi r^2}.
    \label{eq:pt_effective_source}
\end{equation}
Clearly, as $p_r\neq p_t$,
this anisotropy in the pressures confirms what is expected for a radial spinorial/topological hair. This provides a semiclassical gravitational support for the \(q\)-deformed geometry.

\begin{figure}[htb!]
    \centering
    \IfFileExists{fig_energy_conditions_epjc.pdf}{%
    \includegraphics[width=0.96\textwidth]{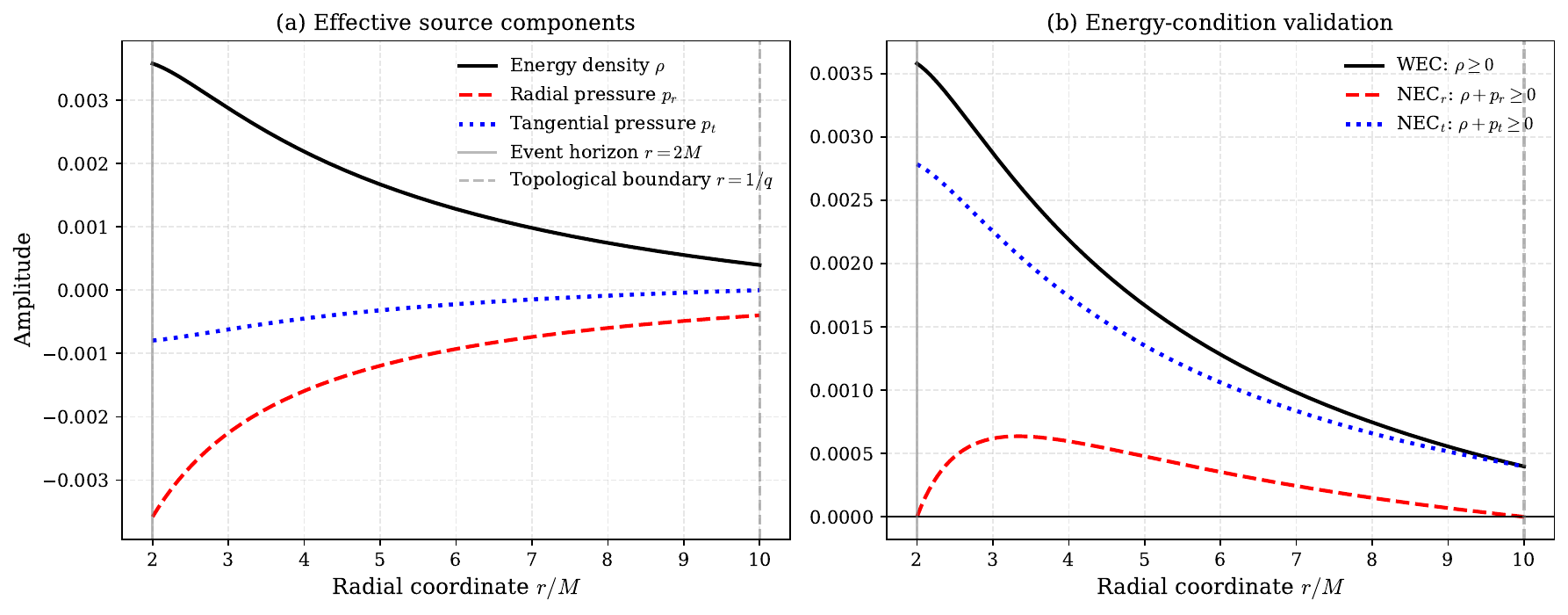}%
    }{%
    \fbox{\parbox[c][5.2cm][c]{0.92\textwidth}{\centering Placeholder for \texttt{fig\_energy\_conditions\_epjc.pdf}.\\Upload the Colab output with this exact filename.}}%
    }
    \caption{
    Effective anisotropic source supporting the topological geometry for \(M=1\) and \(q=0.10\). Left panel shows the energy density \(\rho\), radial pressure \(p_r\), and tangential pressure \(p_t\) in the interval \(r_H\leq r\leq r_{\rm topo}\). The source has positive energy density and anisotropic principal pressures, \(p_r\neq p_t\), as expected for a radial spinorial/topological hair. Right panel displays the weak and null energy-condition combinations. The radial null energy condition saturates at the boundaries and remains non-negative inside the physical interval.
    }
    \label{fig:energy_conditions_epjc}
\end{figure}

The relevant energy-condition combinations are,
\begin{equation}
    \rho+p_r
    =
    -\frac{q(r-2M)(qr-1)}{4\pi r^2},
    \label{eq:nec_radial_effective_source}
\end{equation}
and
\begin{equation}
    \rho+p_t
    =
    \frac{q}{8\pi r^2}
    \left[
        4r-4M+4Mqr-3qr^2
        +(r-M)(qr-1)
    \right].
    \label{eq:nec_tangential_effective_source}
\end{equation}
In the physical exterior region,
\begin{equation}
    2M<r<\frac{1}{q},
\end{equation}
one has \(r-2M>0\) and \(qr-1<0\). Therefore, the radial null energy condition is satisfied inside the interval,
\begin{equation}
    \rho+p_r>0
\end{equation}
and saturates at both boundaries. Also, at the Schwarzschild horizon we have,
\begin{equation}
    p_r(r_H)=-\rho(r_H),
\end{equation}
so that the effective source behaves locally as a radial vacuum tension. This is consistent with the interpretation that the topological hair modifies the radial sector of the geometry without introducing an additional Killing horizon. In Fig. \ref{fig:energy_conditions_epjc} we show the energy density and the radial and tangential pressures as well as the weak and null energy conditions for the physical interval of the radial coordinate of the solution. 

At the effective-field-theory level, this source may be regarded as arising from an action of the schematic form,
\begin{equation}
    S_{\rm eff}
    =
    \int d^4x\sqrt{-g}
    \left[
        \frac{R}{16\pi}
        +
        \mathcal L_{\rm topo}
    \right],
    \label{eq:effective_action_topological_source}
\end{equation}
where \(\mathcal L_{\rm topo}\) encodes the exotic spinorial sector and its associated topological hair. A microscopic realization would involve the modified Dirac Lagrangian,
\begin{equation}
    \mathcal L_{\rm Dirac}^{\rm exo}
    =
    \frac{i}{2}
    \left[
        \bar\Psi\tilde\gamma^\mu\tilde\nabla_\mu\Psi
        -
        (\tilde\nabla_\mu\bar\Psi)\tilde\gamma^\mu\Psi
    \right]
    -
    m\bar\Psi\Psi
    +
    \mathcal L_{\rm hair},
    \label{eq:microscopic_exotic_dirac_lagrangian}
\end{equation}
with
\begin{equation}
    \{\tilde\gamma^\mu,\tilde\gamma^\nu\}
    =
    2\tilde g^{\mu\nu}\mathbb I.
\end{equation}
A complete derivation of \(T_{\mu\nu}^{\rm topo}\) from the microscopic exotic spinorial action requires varying the exotic tetrad and the corresponding spin connection. In the present work, we use the reconstructed anisotropic source above as the semiclassical support of the effective topological geometry.

The radial coefficient in Eq.~\eqref{eq:grr_eff}, now interpreted as part of the semiclassical geometry supported by the source~\eqref{eq:anisotropic_source_def}, has two zeros. The first is the usual Schwarzschild horizon, $\rH=2M$. The second occurs at,
\begin{equation}
    \rtopo=\frac{1}{q}.
    \label{eq:rtopo}
\end{equation}
We refer to this surface as the \emph{topological impedance surface}. For it to be located outside the gravitational horizon, one must impose,
\begin{equation}
    \rtopo>\rH
    \quad\Longleftrightarrow\quad
    \frac{1}{q}>2M
    \quad\Longleftrightarrow\quad
    q<\frac{1}{2M}.
    \label{eq:q_condition}
\end{equation}
If Eq.~\eqref{eq:q_condition} is not satisfied, the impedance surface lies inside the Schwarzschild horizon and cannot act as an externally observable propagation barrier.

This point is central to the physical interpretation. The surface $\rtopo$ is not, in general, an independent Killing horizon associated with the temporal Killing vector $\xi=\partial_t$. Indeed, the norm of this Killing vector in the effective geometry is,
\begin{equation}
    \xi^\mu\xi_\mu
    =
    \tilde g_{tt}
    =
    -f(r)
    =
    -\left(1-\frac{2M}{r}\right).
    \label{eq:killing_norm}
\end{equation}
Therefore, $\xi^\mu\xi_\mu$ vanishes only at $r=2M$, not at $r=1/q$, unless both surfaces coincide in the extremal limit.

This mapping directly extends the flat-space framework established in \cite{dosSantosFilho:2026epjc}, where the geometric decomposition of the exotic Clifford algebra was shown to structurally modify the vector and tensorial currents. When transported to a curved spacetime background, the preferred topological direction $b_\mu = \partial_\mu\varphi$ does not merely induce localized current corrections; through the tetrad deformation, it inherently alters the spin connection of the background geometry. Consequently, the flat-space spin-splitting effect is macroscopically upgraded into a gravitational scattering mechanism, locking the microscopic helicity sectors to the global asymptotic boundary conditions of the black hole.

However, this does not mean that $\rtopo$ is irrelevant for the exotic radial dynamics. Since the effective radial coefficient $\tilde g^{rr}$ also vanishes at $r=1/q$, the surface $\rtopo$ acts as a second asymptotic boundary for exotic spinorial propagation. In this precise sense, it behaves as a cosmological-like effective horizon for the radial exotic problem, even though it is not an independent Killing horizon of the full effective metric. The exotic exterior region is therefore analogous to a finite radial interval bounded by two effective horizons,
\begin{equation}
    r_H<r<r_{\rm topo},
\end{equation}
similar in spirit to the static patch between a black-hole horizon and a cosmological horizon.

The critical condition,
\begin{equation}
    r_{\rm topo}=r_H\,,
\end{equation}
corresponds to,
\begin{equation}
    \frac{1}{q}=2M,
\end{equation}
and should be interpreted as an extremal-like limit for the effective exotic radial geometry. In this sense, the condition $1/q=2M$ plays the role of an extremal black-hole limit. As shown below, this same condition also makes the effective exotic temperature vanish.

\subsection{Effective surface gravity and exotic temperature}
\label{subsec:effective_temperature}

The \(q\)-deformed effective radial geometry \eqref{eq:topological_geometry_source} can be written in the form,
\begin{equation}
    d\tilde s^2
    =
    -A(r)dt^2
    +
    \frac{dr^2}{B(r)}
    +
    r^2d\Omega^2,
    \label{eq:effective_AB_metric}
\end{equation}
with
\begin{equation}
    A(r)=f(r),
    \qquad
    B(r)=f(r)(1-qr)^2.
    \label{eq:AB_functions}
\end{equation}

Because the effective metric has asymmetric radial and temporal coefficients, $A(r)\neq B(r)$, the corresponding tortoise coordinate cannot be obtained by a naive Schwarzschild replacement. It must be derived from the full radial sector of the effective metric.

For radial null propagation, $d\tilde s^2=0$ and $d\Omega^2=0$, Eq.~\eqref{eq:effective_AB_metric} gives,
\begin{equation}
    0
    =
    -A(r)dt^2
    +
    \frac{dr^2}{B(r)}.
\end{equation}
Hence,
\begin{equation}
    \frac{dt}{dr}
    =
    \pm
    \frac{1}{\sqrt{A(r)B(r)}}.
\end{equation}
In order for radial null rays to be written as,
\begin{equation}
    t\pm r_*=\mathrm{constant},
\end{equation}
the appropriate tortoise coordinate must satisfy,
\begin{equation}
    \frac{dr_*}{dr}
    =
    \frac{1}{\sqrt{A(r)B(r)}}.
    \label{eq:tortoise_general_AB}
\end{equation}
For the exotic effective geometry,
\begin{equation}
    A(r)B(r)
    =
    f(r)^2(1-qr)^2.
\end{equation}
Then, as in the exterior exotic region $r_H<r<r_{\rm topo}$, one has $1-qr>0$, we have,
\begin{equation}
    \frac{dr_*}{dr}
    =
    \frac{1}{f(r)(1-qr)},
    \qquad
    \frac{d}{dr_*}
    =
    f(r)(1-qr)\frac{d}{dr}.
    \label{eq:tortoise_topological_derived}
\end{equation}
This shows explicitly that the tortoise coordinate follows from the full effective radial metric sector and not from a direct substitution in the Schwarzschild expression.

Using,
\begin{equation}
    f(r)=1-\frac{2M}{r}
    =
    \frac{r-2M}{r},
\end{equation}
Eq.~\eqref{eq:tortoise_topological_derived} may be integrated explicitly,
\begin{equation}
    r_*(r)
    =
    \int
    \frac{dr}{f(r)(1-qr)}
    =
    \int
    \frac{r\,dr}{(r-2M)(1-qr)}.
\end{equation}
For $q\neq0$ and $q\neq1/(2M)$, partial fraction decomposition gives,
\begin{equation}
    r_*(r)
    =
    \frac{1}{1-2Mq}
    \left[
    2M\ln\left(r-2M\right)
    -
    \frac{1}{q}\ln\left(1-qr\right)
    \right]
    +
    C,
    \label{eq:tortoise_analytic}
\end{equation}
where $C$ is an arbitrary additive constant. In the physical exotic domain,
\begin{equation}
    2M<r<\frac{1}{q},
\end{equation}
both logarithms are real.

Equation~\eqref{eq:tortoise_analytic} shows that,
\begin{equation}\label{tor1}
    r\to r_H^+
    \quad
    \Longrightarrow
    \quad
    r_*\to-\infty,
\end{equation}
whereas,
\begin{equation}\label{tor2}
    r\to r_{\rm topo}^-
    \quad
    \Longrightarrow
    \quad
    r_*\to+\infty.
\end{equation}
Thus, although the coordinate interval $r_H<r<r_{\rm topo}$ is finite in $r$, it becomes an infinite open interval in the tortoise coordinate,
\begin{equation}
    r_*\in(-\infty,+\infty).
\end{equation}
This is precisely why $\rtopo$ plays the role of a cosmological-like effective horizon for the exotic radial wave equation.

For a static and spherically symmetric metric of the form~\eqref{eq:effective_AB_metric}, the surface gravity at a simple Killing horizon $r=r_H$ is,
\begin{equation}
    \kappa_{\rm exo}
    =
    \frac{1}{2}
    \sqrt{A'(r_H)B'(r_H)}.
    \label{eq:surface_gravity_general}
\end{equation}
Since $r_H=2M$ and $f(r_H)=0$, one has,
\begin{equation}
    A'(r_H)=f'(r_H),
    \qquad
    B'(r_H)=f'(r_H)(1-qr_H)^2.
\end{equation}
Therefore,
\begin{equation}
    \kappa_{\rm exo}
    =
    \frac{1}{2}f'(r_H)|1-qr_H|.
    \label{eq:kappa_exotic_general}
\end{equation}
Using
\begin{equation}
    f'(r_H)=\frac{1}{2M},
    \qquad
    r_H=2M,
\end{equation}
we obtain
\begin{equation}
    \kappa_{\rm exo}
    =
    \frac{1}{4M}|1-2Mq|.
    \label{eq:kappa_exotic}
\end{equation}
Thus, the corresponding effective temperature associated with the \(q\)-deformed geometry is,
\begin{equation}
    T_{\rm exo}
    =
    \frac{\kappa_{\rm exo}}{2\pi}
    =
    \frac{1}{8\pi M}|1-2Mq|.
    \label{eq:temperature_exotic}
\end{equation}
Notice that in the limit $q\to0$, the usual Schwarzschild temperature is recovered,
\begin{equation}
    T_{\rm exo}
    \longrightarrow
    T_{\rm Schw}
    =
    \frac{1}{8\pi M}.
\end{equation}
While in the critical regime $q\to1/(2M)$, the effective exotic temperature tends to zero. This behavior is consistent with the interpretation that the topological impedance surface approaches the gravitational horizon and progressively suppresses the exotic emission channel. Moreover, this is also compatible with the third law of thermodynamics.

In the semiclassical interpretation adopted here, \(T_{\rm exo}\) is the temperature associated with the \(q\)-deformed geometry itself. This differs from the Schwarzschild value because the topological spinorial hair changes the radial metric coefficient and therefore the surface gravity. 

\subsection{Asymptotic interpretation and domain of validity}

For a constant non-vanishing \(q\), the effective radial coefficient behaves as,
\begin{equation}
    \tilde g^{rr}
    =
    f(r)(1-qr)^2.
\end{equation}
The physical exterior region relevant for the present work is not the limit \(r\to\infty\), but the finite radial interval,
\begin{equation}
    r_H<r<r_{\rm topo},
    \qquad
    r_{\rm topo}=\frac{1}{q}.
\end{equation}
This interval is mapped by the tortoise coordinate into an infinite open domain,
\begin{equation}
    r_*\in(-\infty,+\infty).
\end{equation}
Thus, the topological boundary replaces spatial infinity as the outer asymptotic endpoint of the radial perturbation problem.

The effective source reconstructed above should be understood as a semiclassical description of the spinorial/topological hair supporting the geometry. A complete microscopic model would require deriving the same anisotropic stress-energy tensor from a fully backreacted exotic spinorial action. This remains beyond the scope of the present work. Nevertheless, the reconstruction in Eqs.~\eqref{eq:rho_effective_source}--\eqref{eq:pt_effective_source} shows that the proposed geometry can be supported by a well-defined anisotropic source and does not have to be interpreted as a sector-dependent metric perceived only by one class of spinors.

\section{Birefringent Dispersion and Topological Friction}
\label{sec:friction}

The modified Clifford algebra implies the exotic mass-shell condition,
\begin{equation}
    \tilde g^{\mu\nu}p_\mu p_\nu+m^2=0.
    \label{eq:mass_shell}
\end{equation}
For radial propagation,
\begin{equation}
    p_\mu=(-\omega,p_r,0,0),
\end{equation}
with
\begin{equation}
    \tilde g^{tt}=-f(r)^{-1},
    \qquad
    \tilde g^{rr}=f(r)(1-qr)^2.
\end{equation}
Equation~\eqref{eq:mass_shell} gives,
\begin{equation}
    -f(r)^{-1}\omega^2+f(r)(1-qr)^2p_r^2+m^2=0.
\end{equation}
Solving for the positive-frequency branch yields,
\begin{equation}
    \omega_+
    =
    \sqrt{
    f(r)
    \left[
    m^2+p_r^2f(r)(1-qr)^2
    \right]
    }.
    \label{eq:omega_plus}
\end{equation}
The radial group velocity is given by,
\begin{align}
    v_g
    &=
    \frac{\partial\omega_+}{\partial p_r} \nonumber\\
    &=
    \frac{1}{2\omega_+}
    \frac{\partial}{\partial p_r}
    \left\{f(r)\left[m^2+p_r^2f(r)(1-qr)^2\right]\right\} \nonumber\\
    &=
    \frac{p_rf(r)^2(1-qr)^2}{\omega_+}.
    \label{eq:group_velocity}
\end{align}
Thus, for $q\neq0$,
\begin{equation}
    \lim_{r\to1/q}v_g=0,
    \label{eq:vg_zero}
\end{equation}
provided that $1/q>2M$. This is the kinematic origin of topological friction. The parameter \(q\) controls how strongly the spinorial/topological hair deforms the radial propagation channel. In the limit \(q=0\), the standard Schwarzschild dispersion is recovered. For \(q\neq0\), the radial group velocity is suppressed near \(r_{\rm topo}\), producing an effective birefringence between the undeformed Schwarzschild channel and the topologically deformed spinorial channel.

\section{Exotic Dirac Perturbations in the Effective Geometry}
\label{sec:dirac_perturbations}

Once the effective geometry is known, the next step is to study linear spinorial perturbations propagating on this background. This directly addresses how fermionic modes interact with a metric that already encodes the non-trivial spinorial topology.

We consider the modified Dirac equation,
\begin{equation}
    \left(i\tilde{\gamma}^{\mu}\Dtilde_{\mu}-m\right)\Psi=0,
    \label{eq:modified_dirac}
\end{equation}
where the effective gamma matrices satisfy the Clifford algebra~\eqref{eq:mod_clifford} and, for the metric \eqref{eq:effective_AB_metric}, are given by,
\begin{equation}
    \tilde\gamma^t
    =
    \frac{\gamma^0}{\sqrt{A(r)}},
    \qquad
    \tilde\gamma^r
    =
    \sqrt{B(r)}\,\gamma^1,
    \qquad
    \tilde\gamma^\theta
    =
    \frac{\gamma^2}{r},
    \qquad
    \tilde\gamma^\phi
    =
    \frac{\gamma^3}{r\sin\theta}.
\end{equation}
Such that the massless Dirac operator can be written as,
\begin{equation}
    i\frac{\gamma^0}{\sqrt{A}}\partial_t\Psi
    +
    i\gamma^1\sqrt{B}
    \left(
        \partial_r
        +
        \frac{A'}{4A}
        +
        \frac{1}{r}
    \right)\Psi
    +
    \frac{i}{r}\slashed{D}_{S^2}\Psi
    =
    0.
\end{equation}
In the following, we focus on massless spinorial perturbations, since they provide the cleanest diagnostic of the topological deformation and are directly relevant for Hawking emission.

\subsection{Separation of variables and effective tortoise coordinate}

For a static and spherically symmetric effective geometry in the radial sector, the radial part of the massless Dirac equation can be written in terms of two first-order coupled equations. The topological deformation enters through the radial derivative. The effective tortoise coordinate was derived in Section~\ref{subsec:effective_temperature} from the full metric functions $A(r)$ and $B(r)$, rather than by a naive Schwarzschild replacement. In the exterior exotic domain,
\begin{equation}
    \rH<r<\rtopo,
    \qquad
    \rtopo=\frac{1}{q},
\end{equation}
where $1-qr>0$, it is given by,
\begin{equation}
    \frac{d}{dr_*}
    =
    f(r)(1-qr)\frac{d}{dr},
    \qquad
    dr_*=
    \frac{dr}{f(r)(1-qr)}.
    \label{eq:tortoise_topological}
\end{equation}
The corresponding analytic expression is Eq.~\eqref{eq:tortoise_analytic}. As stated before, $r_*$ maps the finite radial interval $r_H<r<r_{\rm topo}$ into the infinite interval $-\infty<r_*<+\infty$. This confirms that the topological impedance surface acts as the outer asymptotic boundary of the exotic radial problem, in analogy with a cosmological horizon.

After separating the angular variables in terms of spinor spherical harmonics and applying a suitable rescaling of the radial functions, $\Psi = \frac{\psi}{r\,A^{1/4}}$, the radial problem can be written as the supersymmetric pair,
\begin{equation}
    \left(\frac{d}{dr_*}+W\right)\psi_+=i\omega\psi_-,
    \qquad
    \left(\frac{d}{dr_*}-W\right)\psi_-=i\omega\psi_+,
    \label{eq:first_order_dirac_pair}
\end{equation}
where the angular superpotential keeps the standard massless Schwarzschild form
\begin{equation}
    W(r)
    =
    \frac{|\kappa|}{r}\sqrt{f(r)}.
    \label{eq:superpotential_standard}
\end{equation}
Here $|\kappa|=1,2,3,\ldots$ is the angular spinorial quantum number. The topological deformation does not multiply $W(r)$ directly. Instead, it enters through the effective radial derivative induced by,
\begin{equation}
    \tilde g^{rr}=f(r)(1-qr)^2,
\end{equation}
namely through Eq.~\eqref{eq:tortoise_topological}.

Decoupling Eq.~\eqref{eq:first_order_dirac_pair}, one obtains two Schr\"odinger-like equations,
\begin{equation}
    \frac{d^2\psi_\pm}{dr_*^2}
    +
    \left[
    \omega^2
    -
    V_\pm(r;q)
    \right]\psi_\pm
    =0,
    \label{eq:schrodinger_dirac}
\end{equation}
with superpartner potentials,
\begin{equation}
    V_\pm(r;q)
    =
    W(r)^2
    \pm
    \frac{dW(r)}{dr_*}.
    \label{eq:partner_potentials}
\end{equation}
Since,
\begin{equation}
    \frac{dW}{dr_*}
    =
    f(r)(1-qr)\frac{dW}{dr},
    \label{eq:dWdrstar_topological}
\end{equation}
the topological factor modifies the radial propagation through the tortoise derivative while preserving the standard massless spinorial superpotential. Therefore,
\begin{equation}
    V_\pm(r;q)
    =
    W(r)^2
    \pm
    f(r)(1-qr)\frac{dW}{dr}.
    \label{eq:topological_partner_potentials}
\end{equation}

For explicit use in the numerical engine, the ordinary radial derivative is,
\begin{equation}
    \frac{dW}{dr}
    =
    |\kappa|
    \left[
    \frac{f'(r)}{2r\sqrt{f(r)}}
    -
    \frac{\sqrt{f(r)}}{r^2}
    \right].
    \label{eq:dWdr_explicit}
\end{equation}
Substituting Eq.~\eqref{eq:dWdr_explicit} into Eq.~\eqref{eq:topological_partner_potentials}, the explicit potentials implemented in the numerical engine are,
\begin{equation}
    V_\pm(r;q)
    =
    \frac{\kappa^2 f(r)}{r^2}
    \pm
    |\kappa|f(r)(1-qr)
    \left[
    \frac{f'(r)}{2r\sqrt{f(r)}}
    -
    \frac{\sqrt{f(r)}}{r^2}
    \right].
    \label{eq:explicit_topological_potential}
\end{equation}
In the limit $q \rightarrow 0$, Eq. (75) reduces to the standard massless Dirac potential in Schwarzschild spacetime \cite{Cho:2003qe}. The numerical engine enforces this analytical check symbolically as,
\begin{equation}
    V_\pm(r;q=0)-V_{\rm Cho}(r)=0.
    \label{eq:cho_validation}
\end{equation}
Thus, the topological correction modifies the radial operator $d/dr_*$ rather than the angular superpotential itself.

\subsection{Toward the angular sector: preferred direction and helicity splitting}
\label{subsec:angular_sector_opening}

The radial effective model developed above captures the first and most direct consequence of the exotic spinorial deformation: the modification of the radial metric coefficient,
\begin{equation}
    \tilde g^{rr}=f(r)(1-qr)^2,
\end{equation}
and, consequently, the emergence of the topological impedance surface at
\begin{equation}
    r_{\rm topo}=\frac{1}{q}.
\end{equation}
In this minimal description, the angular spinorial operator is kept in its standard spherical form, so that the superpotential is,
\begin{equation}
    W_{\rm min}(r)
    =
    \frac{|\kappa|\sqrt{f(r)}}{r}.
\end{equation}
The topological information then enters through the tortoise derivative,
\begin{equation}
    \frac{d}{dr_*}
    =
    f(r)(1-qr)\frac{d}{dr}.
\end{equation}

However, this minimal radial truncation does not exhaust the full exotic spinorial structure. In the geometrization of topology program, the topological gradient enters at the level of the Dirac matrices themselves. Therefore, the background,
\begin{equation}
    b_\mu\equiv\partial_\mu\varphi
\end{equation}
should not be viewed only as a scalar deformation of the radial geometry. It also defines a preferred direction for the spinorial dynamics. In this sense, the spinorial/topological sector behaves as an effective Lorentz-violating medium: scalar geometric quantities may remain spherically symmetric, while spinorial quantities can still distinguish helicity and angular-momentum sectors.

This point becomes especially clear from the spin-orbit structure induced by the exotic background. In the non-relativistic limit of the exotic Dirac operator, one expects a Rashba-like coupling of the schematic form,
\begin{equation}
    H_{\rm SO}^{\rm topo}
    \propto
    \left(
        \boldsymbol{\nabla}\varphi
        \times
        \mathbf p
    \right)\cdot\boldsymbol{\sigma}.
    \label{eq:topological_spin_orbit_opening}
\end{equation}
For the radial topological background considered here,
\begin{equation}
    \boldsymbol{\nabla}\varphi
    =
    q\,\hat{\mathbf r},
\end{equation}
one obtains,
\begin{equation}
    \boldsymbol{\nabla}\varphi\times\mathbf p
    =
    q\,\hat{\mathbf r}\times\mathbf p.
\end{equation}
Since,
\begin{equation}
    \mathbf L
    =
    \mathbf r\times\mathbf p
    =
    r\,\hat{\mathbf r}\times\mathbf p,
\end{equation}
this gives,
\begin{equation}
    \hat{\mathbf r}\times\mathbf p
    =
    \frac{\mathbf L}{r},
\end{equation}
and therefore,
\begin{equation}
    H_{\rm SO}^{\rm topo}
    \propto
    \frac{q}{r}
    \mathbf L\cdot\boldsymbol{\sigma}.
    \label{eq:LdotS_topological}
\end{equation}
Thus, even a purely radial topological gradient may affect the angular sector of the spinorial perturbation. The reason is that the preferred radial direction couples to the angular momentum through the spin degrees of freedom.

This observation suggests that the standard angular eigenvalue $|\kappa|$ appearing in the minimal superpotential may be replaced, in the full exotic spinorial theory, by an effective topology- and helicity-dependent angular quantity,
\begin{equation}
    |\kappa|
    \longrightarrow
    \Lambda_{\kappa s}(r,q),
    \qquad
    s=\pm,
    \label{eq:kappa_effective_replacement}
\end{equation}
where $s$ labels helicity-like sectors. The corresponding full exotic superpotential would take the form,
\begin{equation}
    W_{\rm full}^{(s)}(r;q)
    =
    \frac{\sqrt{f(r)}}{r}
    \Lambda_{\kappa s}(r,q),
    \label{eq:full_superpotential_angular}
\end{equation}
leading to helicity-dependent superpartner potentials,
\begin{equation}
    V_\pm^{(s)}(r;q)
    =
    \left[
        W_{\rm full}^{(s)}(r;q)
    \right]^2
    \pm
    \frac{dW_{\rm full}^{(s)}}{dr_*}.
    \label{eq:full_angular_potentials}
\end{equation}
The minimal radial model is recovered when
\begin{equation}
    \Lambda_{\kappa s}(r,q)=|\kappa|.
\end{equation}

A perturbative parametrization of the full angular deformation may be written as,
\begin{equation}
    \Lambda_{\kappa s}(r,q)
    =
    |\kappa|
    +
    s\,\delta\Lambda_\kappa(r,q)
    +
    {\cal O}(q^2),
    \label{eq:Lambda_perturbative_main}
\end{equation}
where the correction $\delta\Lambda_\kappa$ encodes the helicity-dependent response of the spinorial angular operator to the preferred topological direction. Such a deformation would naturally generate birefringent propagation,
\begin{equation}
    V_\pm^{(+)}(r;q)
    \neq
    V_\pm^{(-)}(r;q),
\end{equation}
and, therefore, different greybody factors and damping rates for different helicity sectors.

A leading-order estimate of this angular deformation can be obtained without committing to the full separation of the exotic Dirac operator. Projecting the spin-orbit structure in Eq.~\eqref{eq:LdotS_topological} onto the standard spinor spherical harmonics, one uses the angular identity,
\begin{equation}
    \left(\boldsymbol{\sigma}\cdot\mathbf L+1\right)
    \Omega_{\kappa m}
    =
    -\kappa\,\Omega_{\kappa m},
    \label{eq:sigmaL_spinor_harmonics_main}
\end{equation}
which implies,
\begin{equation}
    \boldsymbol{\sigma}\cdot\mathbf L\,\Omega_{\kappa m}
    =
    -(\kappa+1)\Omega_{\kappa m}.
    \label{eq:LdotS_projection_main}
\end{equation}
This provides the leading-order realization of Eq.~\eqref{eq:Lambda_perturbative_main},
\begin{equation}
    \delta\Lambda_\kappa(r,q)=\eta_\kappa(r)q+{\cal O}(q^2),
\end{equation}
so that,

\begin{equation}
    \Lambda_{\kappa s}(r,q)
    =
    |\kappa|
    +
    s\,\eta_\kappa(r)\,q
    +
    {\cal O}(q^2),
    \label{eq:Lambda_first_order_projection_main}
\end{equation}
where \(s=\pm\) labels helicity-like sectors and \(\eta_\kappa(r)\) absorbs the normalization of the projected exotic angular operator. Equivalently, one may view \(\eta_\kappa(r)\) as the effective radial weight multiplying the eigenvalue of \(\boldsymbol{\sigma}\cdot\mathbf L\) after the full spin connection has been projected onto the angular basis.

Equation~\eqref{eq:Lambda_first_order_projection_main} should not be interpreted as the exact angular spectrum of the full exotic Dirac operator. Rather, it is a controlled structural estimate showing how the radial topological gradient can feed into the angular spinorial sector. The coefficient \(\eta_\kappa(r)\) can only be fixed unambiguously after separating the complete exotic Dirac operator, including the modified Clifford structure and the induced spin connection.

The present work focuses on the minimal radial model, because it is the sector directly fixed by the effective metric component $\tilde g^{rr}$ and by the corrected tortoise coordinate. Nevertheless, Eq.~\eqref{eq:full_superpotential_angular} indicates the natural extension of the formalism. In particular, the non-zero asymptotic plateau found in the minimal model,
\begin{equation}
    V_\infty^{\rm min}
    =
    \kappa^2q^2(1-2Mq),
\end{equation}
would become, in the helicity-dependent theory,
\begin{equation}
    V_\infty^{(s)}
    =
    q^2(1-2Mq)
    \left[
        \Lambda_{\kappa s}(r_{\rm topo},q)
    \right]^2.
    \label{eq:Vinf_helicity_dependent_main}
\end{equation}
Hence, the asymptotic structure of the exotic spinorial potential is not purely radial: it may also depend on the full angular spinorial operator. This is the sector in which the Lorentz-violating and Rashba-like features of exotic spinors are expected to appear most explicitly.

A complete treatment of Eq.~\eqref{eq:kappa_effective_replacement} requires separating the full exotic Dirac operator in the curved black-hole background, including the modified Clifford structure, the induced spin connection, and the preferred-direction spin couplings. We leave this derivation for a subsequent analysis. The role of the present subsection is to make clear that the minimal radial model is not the endpoint of the construction, but the first controlled truncation of a broader helicity-dependent exotic spinorial problem. 
It is worth noting that this emergent helicity-dependent scattering mechanism shares structural similarities with the gravitational sector of the Standard-Model Extension (SME) \cite{Kostelecky:2003fs, Colladay:1998fq} and its associated gauge and fermionic expansions \cite{Casana:2009uq}. In scenarios where local Lorentz invariance is spontaneously broken, a background vector field naturally couples to the spin degrees of freedom, inducing birefringent propagation and non-minimal topological effects \cite{Kostelecky:2008in, Belich:2005wd}. In our framework, however, the Lorentz-violating medium is not an external arbitrary tensor, but is fundamentally rooted in the geometrization of the exotic spinorial topology itself \cite{HoffdaSilva:2022xx, Ahluwalia:2004sz}. This provides a direct topological mechanism for the birefringent features observed in the quasinormal spectrum.

\section{Numerical Diagnostics: Potential Barriers and Quasinormal Modes}
\label{sec:numerical_diagnostics}

At this stage the construction becomes testable. Once the effective radial geometry has been fixed, the exotic Dirac equation no longer provides only a formal deformation of Schwarzschild propagation, it defines a concrete scattering problem. The effective potential determines how exotic spinorial waves are reflected, transmitted, and damped and, therefore, gives a numerical window into the role of the topological impedance surface.

The numerical engine implements Eq.~\eqref{eq:schrodinger_dirac} with the topological deformation entering through the radial operator,
\begin{equation}
    \frac{d}{dr_*}=f(r)(1-qr)\frac{d}{dr}.
\end{equation}
Before solving the topological quasinormal problem, the symbolic validation condition, Eq. \eqref{eq:cho_validation}, 
is imposed. This check ensures that the standard massless Schwarzschild--Dirac potential is recovered when the topological gradient is switched off.

We work in natural units and set $M=1$, so that the Schwarzschild horizon is located at $\rH=2$. For $q>0$, the external exotic domain is finite in the radial coordinate,
\begin{equation}
    r\in(\rH,\rtopo),
    \qquad
    \rtopo=\frac{1}{q},
\end{equation}
and, as analyzed in Eqs. \eqref{tor1} and \eqref{tor2}, this finite radial interval is mapped into an infinite tortoise interval, $r_*\in(-\infty,+\infty)$.

Regarding the numerical evolution, the quasinormal problem should be formulated on a sufficiently large but finite numerical interval in tortoise coordinate,
\begin{equation}
    r_*\in[-L,L],
\end{equation}
with $L$ chosen large enough to approximate the two asymptotic boundaries. In the direct time-domain analysis below we explore both moderate and near-critical values of the topological gradient, $q=0.10,$ $0.20,$ $0.35,$ $0.40,$ $0.45,$ $0.47,$ $0.49$,
which satisfy the condition $q<1/(2M)=0.5$ for $M=1$. The case $q=0$ is kept as an analytical benchmark for the potential, while the finite-patch time-domain analysis focuses on $q>0$. This extended scan is designed to test whether the near-critical regime approaches a stable quasiresonant limit or develops a dynamical instability, as occurs in other modified gravity scenarios \cite{Abdalla:2018cmx}.

\subsection{Minimal radial potential in tortoise coordinate}

The frequency-domain formulation and the time-domain diagnostics reported below use the minimal radial effective model. In this model the topological deformation modifies the radial coefficient of the effective metric and, therefore, the tortoise map, while the spinorial superpotential remains as,
\begin{equation}
    W_{\rm min}(r)
    =
    \frac{|\kappa|\sqrt{f(r)}}{r}.
\end{equation}
Consequently,
\begin{equation}
    V_\pm^{\rm min}(r;q) =
    W_{\rm min}^2 \pm
    \frac{dW_{\rm min}}{dr_*}.
\end{equation}
The same potential can be represented directly as a function of the tortoise coordinate by solving $r=r(r_*)$ in the interval $\rH<r<\rtopo$. All numerical results reported in this section and in Table~\ref{tab:time_domain_damping_kappa1} are obtained within this minimal radial model for which $\Lambda_{\kappa s}(r,q)=|\kappa|$. The helicity-dependent replacement discussed in Section~\ref{subsec:angular_sector_opening} identifies the next analytical layer of the construction and is not used in the present numerical tables.

\begin{figure}[htb!]
    \centering
    \includegraphics[width=0.9\textwidth]{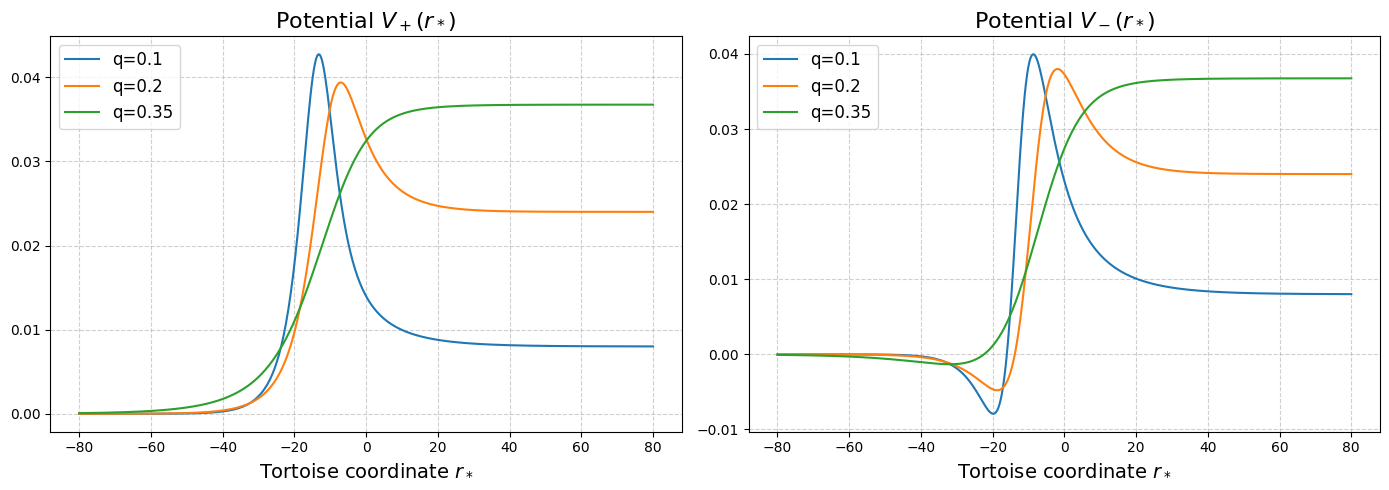}
    \caption{Minimal radial effective superpartner potentials $V_+(r_*)$ (left) and $V_-(r_*)$ (right) for $M=1$, $|\kappa|=1$, and $q=0.10,0.20,0.35$. The left asymptotic region corresponds to the Schwarzschild horizon, where $V_\pm\to0$, while the right asymptotic region corresponds to the topological boundary, where the minimal model approaches the non-zero plateau $V_\infty=\kappa^2q^2(1-2Mq)$. Both potentials display a peak, while only $V_-$ develops a negative well outside the event horizon.}
    \label{fig:potential_rstar_minimal_kappa1}
\end{figure}

Figure~\ref{fig:potential_rstar_minimal_kappa1} shows both superpartner potentials $V_\pm(r_*)$ for $M=1$, $|\kappa|=1$, and selected values of $q$. The left asymptotic region corresponds to the Schwarzschild horizon and satisfies $V_\pm(r_*\to-\infty)=0$. The right asymptotic region corresponds to the topological boundary. In the minimal radial model, this boundary is not a free asymptotic region because the potential approaches the non-zero plateau,
\begin{equation}
    V_\infty
    =
    \kappa^2 q^2(1-2Mq).
    \label{eq:V_infinity_main}
\end{equation}
This behavior follows from the fact that $A(\rtopo)=f(\rtopo)\neq0$ away from the critical limit, while the derivative term $dW/dr_*$ vanishes at $\rtopo$.

The presence of the plateau modifies the outgoing condition at the topological boundary. Instead of the free asymptotic form $e^{+i\omega r_*}$, the appropriate right-boundary behavior in the minimal radial model is,
\begin{equation}
    \psi_\pm
    \sim
    e^{+ik_{\rm topo}r_*},
    \qquad
    k_{\rm topo}
    =
    \sqrt{\omega^2-V_\infty}.
    \label{eq:k_topo_main}
\end{equation}
This point is important for interpreting the frequency-domain shooting results and for constructing stable time-domain simulations.

Moreover, although the asymptotic behavior of both superpartner potentials is the same, we notice that $V_-$ develops a negative minimum before the peak. This negative well could trigger the appearance of unstable modes, however, our results showed no instability for the model. 

\subsection{Asymptotic plateau and oscillatory tail}
\label{subsec:plateau_oscillatory_tail}

The non-zero asymptotic plateau in Eq.~\eqref{eq:V_infinity_main} is not merely a numerical detail. In the minimal radial model it plays the role of an effective asymptotic mass scale for the spinorial perturbation. One may define,
\begin{equation}
    \mu_{\rm eff}
    \equiv V_\infty ^{1/2} = 
    |\kappa|q\sqrt{1-2Mq}.
    \label{eq:mu_eff_plateau}
\end{equation}
Thus, the right asymptotic region behaves as a dispersive medium rather than a free massless channel. This observation provides a natural interpretation for the oscillatory late-time tail observed in the time-domain signal: it may be associated with the non-zero topological plateau, analogously to the oscillatory tails of massive fields in black-hole backgrounds. In the present work we use this feature only as a qualitative diagnostic of the minimal radial model; a complete asymptotic late-time expansion is left for future analysis.

\subsection{Frequency-domain quasinormal spectrum}
\label{subsec:freq_domain_qnm}

\begin{figure*}[htb!]
    \centering
    \includegraphics[width=1.0\textwidth]{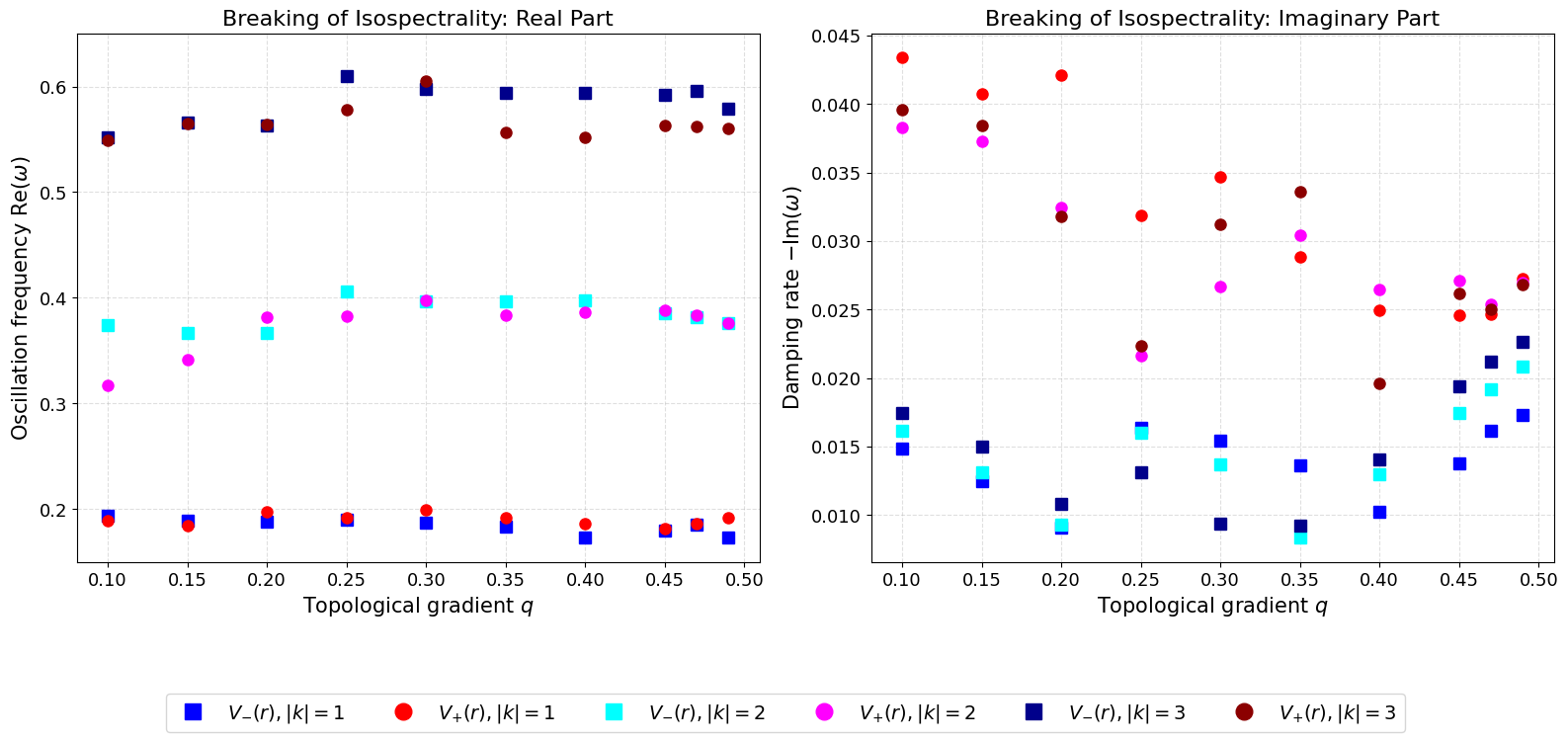}
   \caption{Quasinormal mode spectra for the superpartner potentials $V_{\pm}(r)$ as functions of the topological gradient $q$. Distinct colors indicate the multipoles $|\kappa| = 1, 2, 3$, while squares and circles correspond to $V_-$ and $V_+$, respectively. The left panel shows the oscillation frequencies, which look very similar for the two sectors across the entire $q$-range. The right panel displays the damping rates, where the breaking of isospectrality clearly emerges: $V_+$ consistently exhibits larger damping than $V_-$. This helicity-dependent, birefringent spectral response is a direct signature of the topological hair.}
    \label{fig3_breaking}
\end{figure*}

Now, we consider the quasinormal-mode boundary-value problem associated with exotic spinorial perturbations, defined through the Schr\"odinger-like equation \eqref{eq:schrodinger_dirac}. Near the Schwarzschild horizon, the physical solution is purely ingoing,
\begin{equation}
    \psi_\pm \sim e^{-i\omega r_*}, \qquad r_*\to-\infty,
    \label{eq:qnm_inner_bc}
\end{equation}
whereas near the topological boundary, the minimal radial model imposes the outgoing condition
\begin{equation}
    \psi_\pm \sim e^{+ik_{\rm topo}r_*}, \qquad r_*\to+\infty,
    \label{eq:qnm_outer_bc}
\end{equation}
with $k_{\rm topo}$ given by Eq.~\eqref{eq:k_topo_main}. To extract the exact quasinormal frequencies $\omega = \omega_{\rm re} + i\omega_{\rm im}$, we solve the eigenvalue problem subject to these boundary conditions. Figure~\ref{fig3_breaking} presents the full spectrum as a function of $q$, while Table~\ref{tab:qnm_near_critical} details the near-critical behavior for the multipoles $|\kappa|=1,2,3$, and the time-domain damping rates are summarized in Table~\ref{tab:time_domain_damping_kappa1}.

The numerical results reveal some crucial physical effects, which we can see in Fig. \ref{fig3_breaking}. 
First, the real part of the fundamental frequency ($\omega_{\rm re}$) 
is the lowest for both potentials, and in general, these frequencies 
are only slightly affected by $q$. However, they are clearly shifted 
with respect to the Schwarzschild case. This shift, although small, 
provides a direct dynamical signature of the topological impedance 
surface, modifying the oscillation spectrum of the field. Second, and 
most importantly, this modification manifests itself strongly in a 
helicity-dependent manner: while the real parts of $V_+$ and $V_-$ 
remain nearly degenerate (as seen in the left panel), the imaginary 
parts exhibit a pronounced splitting (right panel). This clear 
separation in the damping rates provides an unmistakable, direct 
signature of the birefringent propagation induced by the topological 
hair.

\begin{figure}[htb!]
    \centering
    \includegraphics[width=0.75\textwidth]{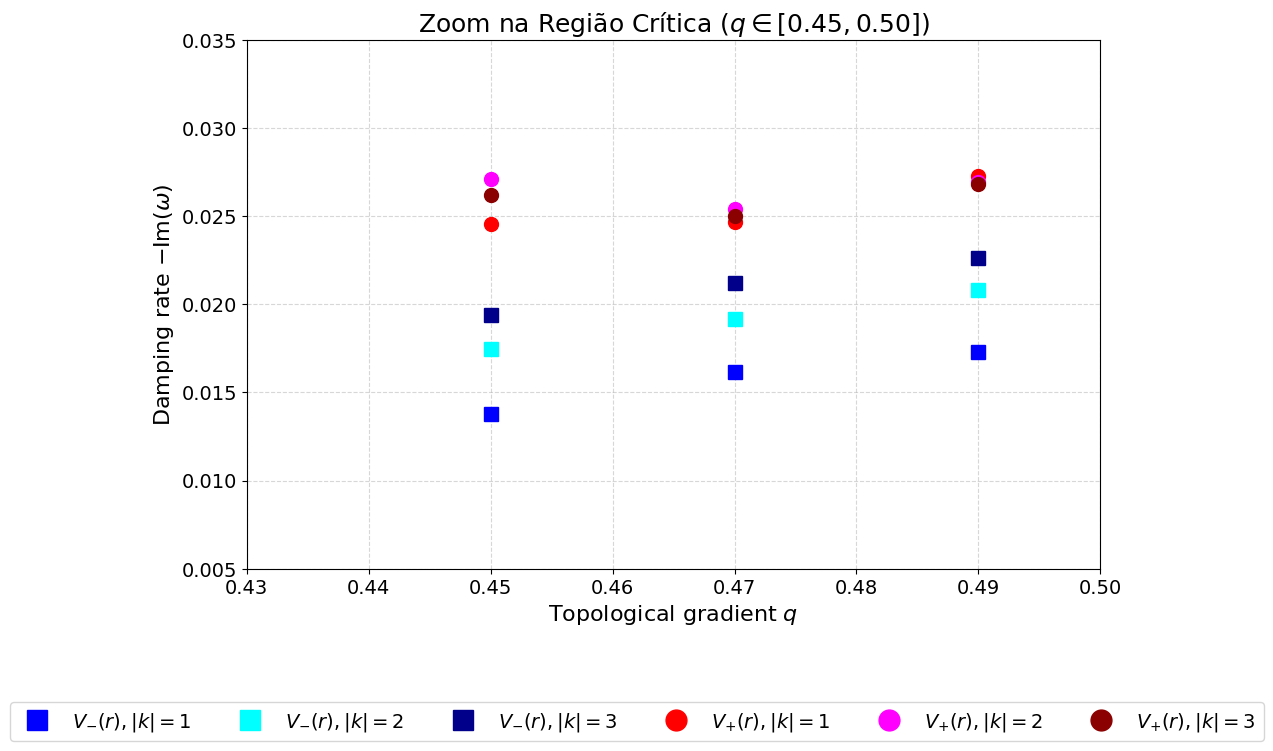}
    \caption{Close-up view of the imaginary part of the quasinormal frequencies, $-\mathrm{Im}(\omega)$, in the interval $0.44 \le q < 0.50$. Squares and circles correspond to $V_-(r)$ and $V_+(r)$, respectively, while colors denote the multipoles $|\kappa|=1,2,3$ as in Fig.~\ref{fig3_breaking}. This magnification clarifies that, although the $V_+$ modes for different multipoles converge towards nearly the same value and cluster together, they remain strictly separated from the $V_-$ modes. Therefore, no physical crossing or numerical overlap occurs between the two superpartner sectors in this region.}
    \label{fig4zoomimaginary}
\end{figure}

Moreover, the imaginary part of the quasinormal frequency, $\mathrm{Im}(\omega)$, 
remains negative for all numerically resolved modes in both potential sectors. 
Consequently, the corresponding perturbations are damped in time, ensuring the 
dynamical stability of the background geometry against these modes. The damping 
behavior, however, differs substantially between the two superpartner potentials.

For the $V_+$ sector, the magnitude of the damping rates exhibits an overall 
decreasing trend as the topological parameter $q$ increases. Remarkably, as the 
system approaches the near-critical regime ($q \to 0.50$), the damping rates for 
different multipoles converge toward a degenerate value. Despite this clustering, 
the damping does not approach zero, indicating that no quasiresonant modes are 
formed within the investigated range. This overall trend is qualitatively 
consistent with the behavior of spinorial perturbations in Schwarzschild--de 
Sitter black holes, where approaching the extremal solution also slows down the 
oscillation, displaying longer damping times~\cite{Zhidenko:2003wq}.

In contrast, the $V_-$ sector exhibits a highly non-monotonic dependence on $q$. 
The magnitude of the damping rates initially decreases, reaching local minima at 
intermediate values of the topological gradient, before growing again as $q$ 
approaches the critical limit. As explicitly detailed in 
Fig.~\ref{fig4zoomimaginary}, the $V_-$ spectrum remains strictly separated from 
the clustered $V_+$ modes in this near-critical regime. This profound separation 
between the damping spectra reflects the different effective geometries probed by 
the two helicity sectors. Such helicity-dependent dynamical response undeniably 
confirms the breaking of isospectrality and provides a robust signature of the 
birefringent propagation induced by the topological hair.

In fact, a fundamental property of standard Schwarzschild black holes is 
the isospectrality between the superpartner potentials $V_+$ and $V_-$, which 
is typically guaranteed by Darboux transformations. However, the spinorial 
topological deformation explicitly breaks this symmetry. As previously 
discussed, the emergence of the non-zero asymptotic plateau $V_\infty$ 
modifies both the boundary conditions and the radial tortoise derivative. 
This structural change induces a negative well in the $V_-(r;q)$ potential 
near the event horizon before it asymptotically converges to the plateau.

To rigorously quantify this effect, we computed the quasinormal frequencies for $V_-$, which governs the dynamics of the opposite helicity sector. The behavior of these modes in the near-critical regime ($q \to 0.50$), where the divergence between the superpartner sectors becomes more pronounced, is detailed in Table~\ref{tab:qnm_near_critical} for multipole numbers $|\kappa|=1,2,3$. Alongside the full spectrum illustrated in Fig.~\ref{fig3_breaking}, which displays the global separation between the $V_+$ and $V_-$ damping rates across the entire parameter space, these numerical results unequivocally confirm the complete breakdown of isospectrality induced by the topological deformation.
This stark divergence in the damping rates implies that fermionic fields with 
opposite helicities interact with the topological hair in fundamentally 
different ways. Ultimately, one helicity channel is more strongly damped than 
its superpartner, being effectively filtered by the geometry. This provides 
the precise dynamical origin of the birefringent Hawking emission.

\begin{table}[htb!]
    \centering
    \caption{
        Quasinormal frequencies in the near-critical regime ($q \to 0.50$) for both superpartner potentials $V_{\pm}(r)$. 
        The data correspond to the magnified view presented in Fig.~\ref{fig4zoomimaginary}. 
        For the $V_+$ sector, the damping rates exhibit a tendency to cluster around a common value as $q$ increases. In contrast, the $V_-$ sector keeps a separation with smaller damping magnitudes. 
        This stark contrast unequivocally confirms the breakdown of isospectrality between the two helicity sectors.
    }
    \label{tab:qnm_near_critical}
    \begin{tabular}{ccccccc}
        \toprule
        & & \multicolumn{2}{c}{$V_+$ Sector} & \multicolumn{2}{c}{$V_-$ Sector} \\
        \cmidrule(lr){3-4} \cmidrule(lr){5-6}
        $q$ & $|\kappa|$ & $\mathrm{Re}(\omega)$ & $\mathrm{Im}(\omega)$ & $\mathrm{Re}(\omega)$ & $\mathrm{Im}(\omega)$ \\
        \midrule
        \multirow{3}{*}{0.45} 
        & 1 & 0.181245 & -0.024572 & 0.179333 & -0.013794 \\
        & 2 & 0.388442 & -0.027095 & 0.385688 & -0.017477 \\
        & 3 & 0.562937 & -0.026195 & 0.592026 & -0.019424 \\
        \midrule
        \multirow{3}{*}{0.47} 
        & 1 & 0.186404 & -0.024658 & 0.185316 & -0.016148 \\
        & 2 & 0.383256 & -0.025399 & 0.381580 & -0.019176 \\
        & 3 & 0.561987 & -0.025009 & 0.595366 & -0.021206 \\
        \midrule
        \multirow{3}{*}{0.49} 
        & 1 & 0.191475 & -0.027290 & 0.173363 & -0.017326 \\
        & 2 & 0.375974 & -0.026952 & 0.376080 & -0.020812 \\
        & 3 & 0.559863 & -0.026840 & 0.578702 & -0.022657 \\
        \bottomrule
    \end{tabular}
\end{table}

To ensure that the observed breaking of isospectrality is not an artifact of the numerical root-finding procedure, we performed a local stability analysis of the quasinormal frequencies. For the specific modes in the near-critical region ($q=0.25$ and $q=0.40$) and for all multipoles $|\kappa|=1,2,3$, we fixed the real part of the frequency, $\mathrm{Re}(\omega)$, to the value obtained by the solver and systematically perturbed the initial guess of the imaginary part, $\mathrm{Im}(\omega)$, within the range $\delta = \pm 0.001, \pm 0.002, \pm 0.005$. In all cases, the extracted imaginary part remained unchanged within numerical precision with a maximum absolute deviation of $\left|\Delta \mathrm{Im}(\omega)\right| \lesssim 10^{-13}$, which is orders of magnitude below the adopted convergence tolerance of $10^{-6}$. This confirms the local numerical stability and uniqueness of the roots used in the analysis.

\subsection{Time-domain ringdown diagnostic}
\label{subsec:time_domain_diagnostic}

As a direct time-domain diagnostic of the same radial problem, we evolve the perturbation directly in the tortoise coordinate. The equation solved numerically is,
\begin{equation}
    \partial_t^2\Psi
    -
    \partial_{r_*}^2\Psi
    +
    V_\pm(r_*)\Psi
    =
    0.
    \label{eq:time_domain_wave}
\end{equation}
The initial data are chosen as a Gaussian packet localized to the left of the potential barrier ($r_{*,0} = -30M$). 
\begin{equation}
    \Psi(0,r_*)
    =
    \exp\left[
    -\frac{(r_*-r_{*,0})^2}{2\sigma^2}
    \right],
    \qquad
    \partial_t\Psi(0,r_*)=0\,.
\end{equation}
Physically, this corresponds to exciting the effective cavity from the inner region near the event horizon. The wavepacket interacts with the barrier from the inside and the transmitted signal is extracted at a fixed observer position outside the peak, chosen here as $r_*^{\rm obs}=0$. Because quasinormal modes are intrinsic resonant properties of the spacetime geometry, their frequencies and damping rates are robustly excited regardless of whether the initial perturbation impinges on the barrier from the left or from the right. 

The numerical domain is a large but finite tortoise-coordinate interval $r_*\in[-L,L]$. Since the exact asymptotic endpoints correspond to $r_*=\pm\infty$, late-time reflections or tail effects may appear if the evolved signal reaches the numerical boundaries. For this reason, the damping is extracted only from the initial ringdown window.

\begin{figure}[htb!] 
    \centering
    \includegraphics[width=0.85\textwidth]{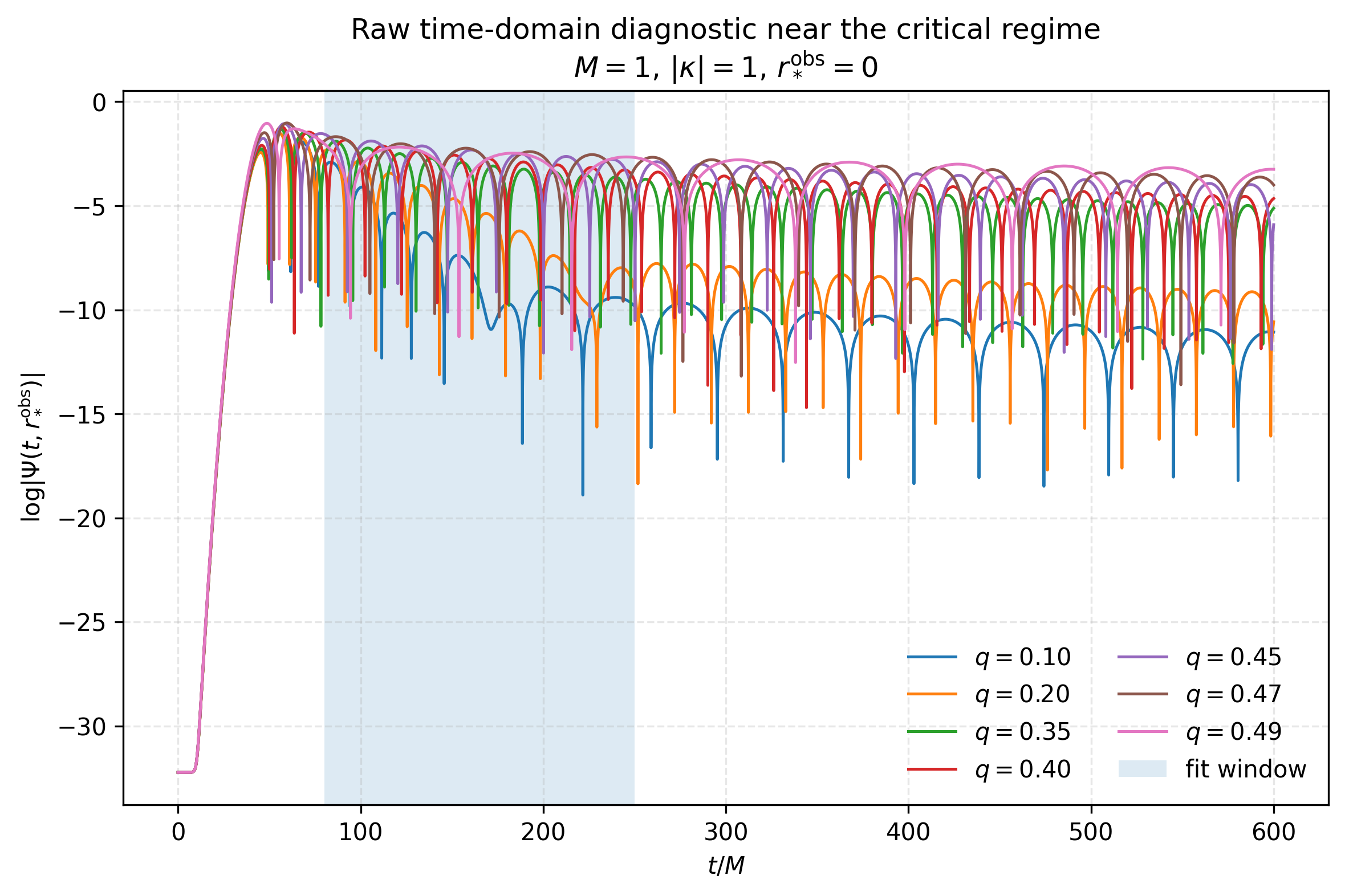}
    \caption{
Time-domain signals $\Psi(t,r_*^{\rm obs})$ for the effective potential $V_+$ with $M=1$ and $|\kappa|=1$. The shaded region $80 \leq t/M \leq 250$ indicates the window used to extract the damping rate. The varying slopes reveal a non-monotonic dependence on $q$: the decay is slower from $q \approx 0.35$ as $q \to 0.50$, and faster for lower $q$ in agreement with the spectrum shown in Fig.~\ref{fig3_breaking}.
}
    \label{fig:time_domain_raw_log_abs_kappa1}
\end{figure}

The raw profile requested in the time domain is the logarithm of the perturbation amplitude, shown in Fig.~\ref{fig:time_domain_raw_log_abs_kappa1}. As the zero crossings of $\Psi$ produce sharp downward spikes in $\log|\Psi|$, the damping rate is extracted more robustly from the Hilbert envelope \cite{Sakai:2017eho},
\begin{equation}
    \mathcal E(t)
    =
    \left|
    \mathrm{Hilbert}[\Psi(t,r_*^{\rm obs})]
    \right|.
\end{equation}
In the ringdown window we fit,
\begin{equation}
    \log\mathcal E(t)
    \simeq
    a-\gamma_{\rm time} t,
    \label{eq:ringdown_fit}
\end{equation}
where $a$ is a constant amplitude offset, so that the extracted damping rate directly correlates with the imaginary part of the frequency-domain spectrum computed in the previous subsection, i.e., 
$\gamma_{\rm time}\simeq-\mathrm{Im}(\omega)$.
The time-domain extraction, therefore, provides a robust, independent validation of the frequency-domain behavior. 

To ensure the robustness of the time-domain integration and prevent contamination of the ringdown signal by artificial boundary reflections, the wave equation was evolved on a uniform tortoise grid $r_*\in[-L,L]$ with spatial resolution $\Delta r_* = 0.05M$ and time step $\Delta t = 0.0225M$, satisfying the Courant--Friedrichs--Lewy condition. Absorbing (Sommerfeld) boundary conditions were implemented at both edges of the computational domain. The extended scan reported in Table~\ref{tab:time_domain_damping_kappa1} adopts $L = 800M$ and the extraction window $80M \leq t \leq 250M$. With this choice reflections from the boundaries can only return to the observation point after a round-trip time of order $2L$, which lies well outside the fitted ringdown interval.

As a direct convergence test, the calculation was repeated for representative near-critical values $q = 0.35, 0.45, 0.49$ by progressively increasing the box size from $L = 800M$ to $L = 1200M$ and $L = 1600M$, while keeping the same spatial resolution and fitting window. The extracted damping rates remained unchanged within the displayed numerical precision, confirming that the observed monotonic behavior of the damping rate is a physical feature of the $V_+$ spectrum, in full agreement with the frequency-domain results shown in Fig.~\ref{fig3_breaking}, and not an artifact of the finite computational domain.
\begin{table}[htb!]
    \centering
   \caption{
    Time-domain damping rates extracted from the Hilbert-envelope fit~\eqref{eq:ringdown_fit} for $V_+$ with $M=1$, $|\kappa|=1$, and $r_*^{\rm obs}=0$. While the time-domain extraction intrinsically carries finite-window transient errors leading to quantitative deviations from the exact asymptotic frequency-domain poles ($\omega_{\rm im}$), it perfectly complements and validates the qualitative trend: the damping rate exhibits a monotonic decrease as the critical regime is approached, remaining strictly positive up to $q=0.49$ and, thus,  ensuring dynamical stability.
}
   
    \label{tab:time_domain_damping_kappa1}
    \begin{tabular}{ccc}
        \toprule
        $q$ & $q/q_{\rm crit}$ & $\gamma_{\rm time}$ \\
        \midrule
        $0.10$ & $0.20$ & $0.04161$ \\
        $0.15$ & $0.30$ & $0.04100$ \\
        $0.20$ & $0.40$ & $0.04064$ \\
        $0.25$ & $0.50$ & $0.03000$ \\
        $0.30$ & $0.60$ & $0.02000$ \\
        $0.35$ & $0.70$ & $0.01079$ \\
        $0.40$ & $0.80$ & $0.00945$ \\
        $0.45$ & $0.90$ & $0.00697$ \\
        $0.47$ & $0.94$ & $0.00563$ \\
        $0.49$ & $0.98$ & $0.00483$ \\
        \bottomrule
    \end{tabular}
\end{table}

\begin{figure}[htb!]
    \centering
    \includegraphics[width=0.76\textwidth]{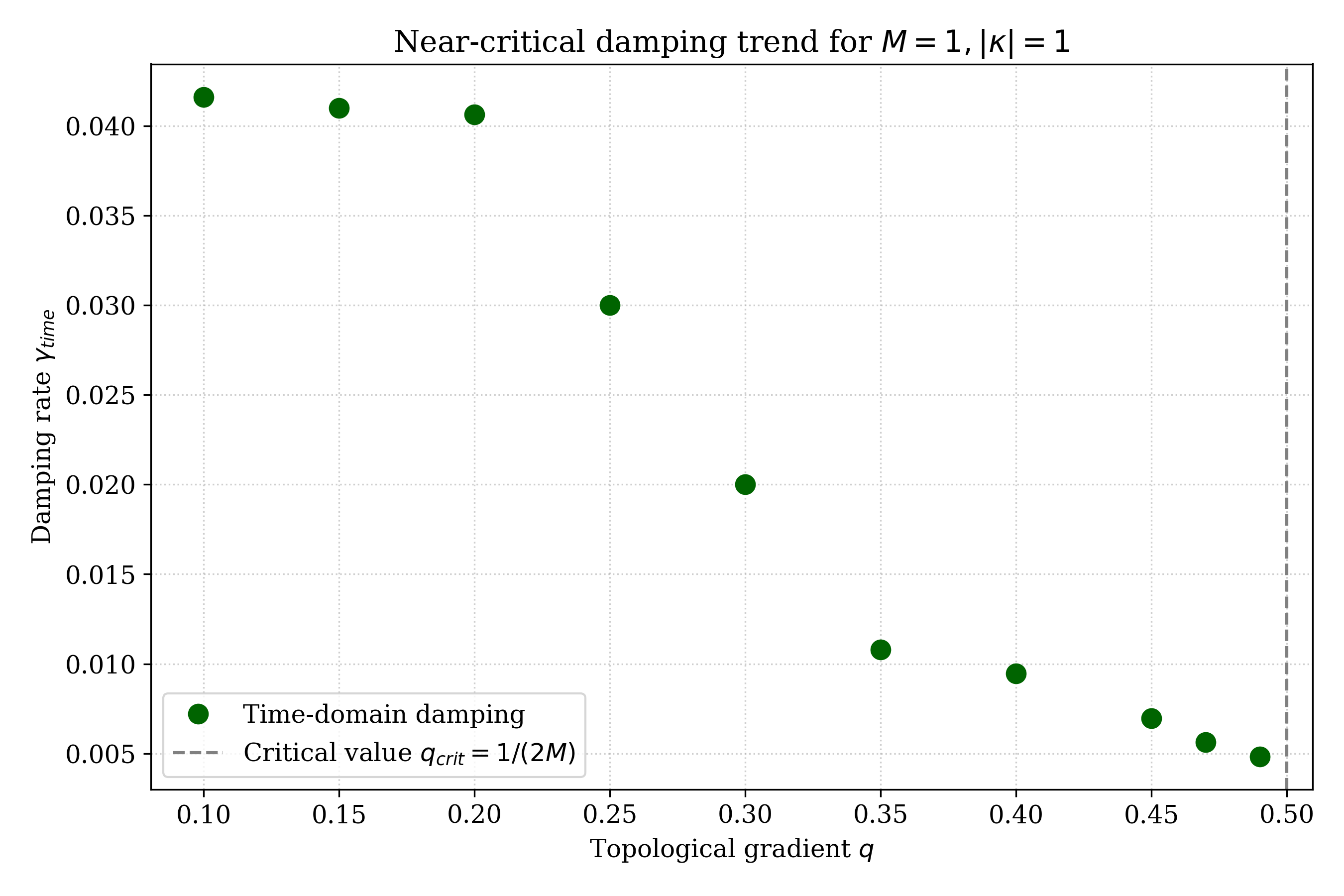}
   \caption{
   Time-domain damping rate $\gamma_{\rm time}$ for the $V_+$ sector shown in Tab. \ref{tab:time_domain_damping_kappa1}. The damping rate decreases monotonically as $q \to q_{\rm crit}=1/(2M)$ and remains strictly positive up to $q=0.49$, ruling out quasiresonant modes, confirming that the $V_+$ sector remains dissipative for all $q$ considered here.
}
    \label{fig:time_domain_damping_vs_q_kappa1}
\end{figure}

Therefore, the time-domain analysis provides independent numerical corroboration of the physical picture inferred from the static effective potential barrier and the exact quasinormal spectrum: an increase in the topological gradient \(q\) progressively reduces the damping rate, thereby prolonging the lifetime of the topologically deformed spinorial perturbation. As shown in Fig.~\ref{fig:time_domain_damping_vs_q_kappa1}, the damping coefficient \(\gamma_{\rm time}\) extracted for the \(V_+\) sector exhibits no change of sign (i.e., it remains strictly positive) up to \(q = 0.49\), which indicates that the near-critical regime remains dynamically stable and does not develop quasiresonant or unstable modes within the explored parameter space.

It should be noted, however, that this numerical result serves as a controlled diagnostic of the minimal radial model, where the angular part of the spinorial field is treated as a fixed eigenmode of the unperturbed angular operator. In the full exotic spinorial model, as discussed in Section~\ref{subsec:angular_sector_opening}, the backreaction of the topological defect on the angular eigenfunctions introduces additional curvature-dependent couplings to the helicity states. These effects may induce helicity-dependent damping rates and a modification of the effective potential barrier beyond the radial approximation, which will be addressed in future work.

\section{Birefringent Hawking Emission as a Greybody-Filter Effect}
\label{sec:hawking}

The analysis performed in Section \ref{subsec:angular_sector_opening} clarifies the role of the topological deformation in the emission process. The topological impedance surface is not an additional Killing horizon, but the \(q\)-deformed geometry possesses a modified radial coefficient and therefore a modified effective surface gravity. As a result, the emission associated with the topologically deformed spinorial channel is shaped by two ingredients: the effective temperature of the deformed geometry and the greybody factor associated with the deformed radial potential.

In the undeformed limit \(q=0\), the emission rate is controlled by the standard Schwarzschild temperature \cite{Hawking:1975vcx, Unruh:1976db, Page:1976df},
\begin{equation}
    T_{\rm Schw}
    =
    \frac{1}{8\pi M}.
\end{equation}
For the topologically deformed spinorial channel, the relevant effective temperature is determined by the surface gravity of the \(q\)-deformed geometry. Therefore, the corresponding emission rate can be written as,
\begin{equation}
    \frac{dP_{\rm exo}^{(s)}}{d\omega}
    =
    \frac{1}{2\pi}
    \frac{\Gamma_{\rm exo}^{(s)}(\omega;q)}
    {e^{\omega/T_{\rm exo}}+1},
    \label{eq:hawking_rate_exotic_helicity}
\end{equation}
where
\begin{equation}
    T_{\rm exo}
    =
    \frac{1}{8\pi M}|1-2Mq|
    \label{eq:hawking_temperature_exotic_section}
\end{equation}
and $\Gamma_{\rm exo}^{(s)}(\omega;q)$ is the greybody factor associated with the topologically deformed spinorial potential for the helicity sector $s=\pm$. The explicit helicity dependence arises because the effective potential $V_\pm^{(s)}$ in Eq.~\eqref{eq:full_angular_potentials} depends on the angular operator $\Lambda_{\kappa s}(r,q)$, which distinguishes the two helicity channels. Thus, while the effective temperature $T_{\rm exo}$ is a global property of the deformed geometry, the greybody factor filters each helicity sector differently, leading to a genuinely birefringent emission spectrum: one helicity is transmitted with higher probability than the other.

In the Schwarzschild limit, the greybody factor is determined by the usual massless spinorial potential. For \(q\neq0\), however, the topological factor in Eq.~\eqref{eq:grr_eff} modifies the radial propagation, the effective potential, and the corresponding transmission coefficient,
\begin{equation}
    \Gamma_{\rm exo}^{(s)}(\omega;q)
    \neq
    \Gamma_{\rm std}^{(s)}(\omega).
\end{equation}
The birefringent character of the emission should, therefore, be understood as the difference between the undeformed Schwarzschild channel and the topologically deformed spinorial channel. The latter is associated with both the modified temperature \(T_{\rm exo}\) and the modified transmission coefficient \(\Gamma_{\rm exo}^{(s)}(\omega;q)\).

\subsection{Origin of the greybody factor}
\label{subsec:origin_greybody}

The greybody factor arises from the scattering problem associated with the radial Eq. \eqref{eq:schrodinger_dirac}.

Near the Schwarzschild horizon, $r\to r_H$, the effective potential vanishes and the physical boundary condition is purely ingoing,
\begin{equation}
    \psi_+
    \sim
    e^{-i\omega r_*}.
\end{equation}
In the outer region, the wave is partially reflected and partially transmitted by the effective potential barrier. Therefore, the greybody factor is identified with the transmission probability through this barrier,
\begin{equation}
    \Gamma_{\rm exo}^{(s)}(\omega;q)
    =
    |T_{\omega q}^{(s)}|^2.
\end{equation}

In the WKB approximation, for frequencies satisfying $\omega^2<V_+^{\rm max}$, the classically forbidden region is bounded by the turning points $r_1$ and $r_2$, defined by,
\begin{equation}
    V_+(r_i;q)=\omega^2,
    \qquad
    i=1,2.
\end{equation}
The WKB action across the barrier is,
\begin{equation}
    S^{(s)}(\omega;q)
    =
    \int_{r_1}^{r_2}
    \sqrt{V_+^{(s)}(r;q)-\omega^2}\,dr_*.
    \label{eq:wkb_action_greybody_helicity}
\end{equation}
Thus, the leading transmission probability becomes,
\begin{equation}
    \Gamma_{\rm exo}^{(s)}(\omega;q)
    \simeq
    \exp[-2S^{(s)}(\omega;q)].
    \label{eq:greybody_wkb_transmission_helicity}
\end{equation}
Equivalently, in a uniform WKB transmission form, one may write,
\begin{equation}
    \Gamma_{\rm exo}^{(s)}(\omega;q)
    \simeq
    \frac{1}
    {1+\exp\left[2S^{(s)}(\omega;q)\right]}.
    \label{eq:greybody_wkb_fermi_form_helicity}
\end{equation}
The topological deformation enters this expression through both the effective potential $V_+^{(s)}(r;q)$ and the tortoise measure \eqref{eq:tortoise_topological}.

An important subtlety arises from the fact that the asymptotic wave numbers at the two boundaries are different. In the standard Schwarzschild case, both ingoing and outgoing modes have the same asymptotic frequency \(\omega\). In the present exotic geometry, however, the outgoing mode at the topological boundary has wave number \(k_{\rm topo} = \sqrt{\omega^2 - V_\infty}\), as given by Eq.~\eqref{eq:k_topo_main}. The conservation of probability flux across the barrier, then, requires that the transmission coefficient be corrected by the ratio of group velocities,
\begin{equation}
    \Gamma_{\rm exo}^{(s)}(\omega;q)
    =
    \frac{v_{\rm in}}{v_{\rm out}}
    |T_{\omega q}^{(s)}|^2,
\end{equation}
where \(v_{\rm in}\) and \(v_{\rm out}\) are the group velocities near the horizon and at the topological boundary, respectively. This correction is precisely the kinematic manifestation of the topological friction discussed in Section~\ref{sec:friction}. For \(\omega^2 < V_\infty\), the outgoing mode becomes evanescent and the transmission probability vanishes identically,
\begin{equation}
    \Gamma_{\rm exo}^{(s)}(\omega;q)
    =
    0,
    \qquad
    \omega \leq \mu_{\rm eff},
\end{equation}
with \(\mu_{\rm eff} \) defined in Eq.~\eqref{eq:mu_eff_plateau}. Thus, the topological impedance surface acts as an effective infrared cutoff for the exotic emission channel, filtering out modes with frequency below the effective mass scale set by the asymptotic plateau.

The QNM results are not themselves Hawking spectra. Rather, they provide an independent diagnostic of the same effective barrier that controls the greybody factor. Therefore, the suppression of the exotic emission channel follows from geometric filtering by the effective potential. This interpretation is compatible with the fact that the exotic emission channel is governed by both the effective temperature \(T_{\rm exo}\) and the greybody factor \(\Gamma_{\rm exo}^{(s)}(\omega;q)\).

Crucially, the same geometric barrier that suppresses the greybody factor is also responsible for the decrease in the damping rate \(\gamma_{\rm time}\) observed in Section~\ref{sec:numerical_diagnostics}. As \(q \to q_{\rm crit}\), the barrier becomes broader and taller, reducing the leakage of the spinorial modes towards the topological boundary. This reduction in leakage manifests in the frequency domain as a smaller imaginary part \(\omega_{\rm im}\) (longer-lived modes) and in the time domain as a smaller \(\gamma_{\rm time}\). The combined effect—strong suppression of the greybody factor and decrease of the damping rate—provides the kinematic and thermodynamic mechanism for the formation of long-lived topological remnant candidates, which we discuss in Section~\ref{sec:remnants}. A complete numerical computation of \(\Gamma_{\rm exo}^{(s)}(\omega;q)\) requires solving the full scattering problem with the plateau boundary condition in Eq.~\eqref{eq:k_topo_main}; here we identify the geometric filtering mechanism and leave the exact greybody spectrum for future work.

\section{Kinematic Segregation and Long-Lived Remnant Candidates}
\label{sec:remnants}

The spinorial perturbation analysis provides a clearer physical picture of the remnant scenario. The undeformed Schwarzschild limit and the topologically deformed spinorial channel do not exhibit the same radial propagation. While the limit \(q=0\) reproduces the usual Schwarzschild potential, the \(q\neq0\) channel is affected by the topological impedance factor, which acts not only as an overall suppressor but also as a birefringent filter segregating the different helicity sectors, as discussed in Sections~\ref{subsec:angular_sector_opening} and~\ref{sec:hawking}.

The critical scale of the exotic geometry is determined by the coincidence between the Schwarzschild horizon and the topological impedance surface,
\begin{equation}
    \rH=\rtopo.
\end{equation}
Since $\rH=2M$ and $\rtopo=1/q$, one obtains the critical mass,
\begin{equation}
    \Mcrit
    =
    \frac{1}{2q}.
    \label{eq:Mcrit_remnant}
\end{equation}

The phenomenological viability of this critical state as an evaporation endpoint depends on the dynamical evolution of the topological hair. If the total topological charge is conserved during evaporation, the effective gradient $q$ is expected to grow as the black hole shrinks. This drives the system toward the critical regime $2Mq \to 1$. As the black hole approaches this limit, the available radial domain for exotic perturbations, $\rH < r < \rtopo$, becomes increasingly compressed, behaving effectively as a narrowing resonant cavity. Simultaneously, the effective exotic temperature,
\begin{equation}
    T_{\rm exo}
    =
    \frac{1}{8\pi M}|1-2Mq|,
\end{equation}
tends to zero. This provides a direct thermodynamic mechanism for quenching the exotic emission channel.

The greybody-filter interpretation adds the necessary kinematic suppression mechanism. As the geometry compresses toward criticality, exotic spinorial modes must tunnel through an increasingly restrictive and helicity-selective effective potential. The time-domain evolution in the tortoise coordinate, summarized in Table~\ref{tab:time_domain_damping_kappa1} and Fig.~\ref{fig:time_domain_damping_vs_q_kappa1}, confirms that the damping rate $\gamma_{\rm time}$ decreases as the topological deformation approaches the critical regime. Trapped between the horizon and the approaching topological boundary, the perturbations bounce within the shrinking effective cavity, resulting in longer-lived exotic modes rather than a near-critical instability. The box-size convergence test validates that this robust trend is physical and not an artifact of numerical boundaries.

This combined framework produces a solid suppression mechanism: as the spinorial/topological hair approaches the critical regime, the effective temperature vanishes and the deformed transmission barrier geometrically blocks the corresponding emission channels with one helicity being suppressed faster than the other. The endpoint of this process should, therefore, be interpreted as a long-lived topological remnant candidate. A definitive proof of its absolute stability would require a fully dynamical evaporation model including backreaction, the time evolution of the topological background, and the competitive drain of all available standard radiative channels.

\section{Final Remarks}
\label{sec:concl}

In this work we extended the geometrization of topology program to the propagation of exotic spinors in a Schwarzschild black-hole background. By incorporating the topological gradient into the tetrad structure, we obtained a \(q\)-deformed effective geometry and reconstructed the anisotropic source required to support it semiclassically. The main result is the emergence of a topological impedance surface at $r=1/q$, where the radial group velocity of exotic spinors is suppressed.

We emphasized that this surface should not be interpreted as an independent Killing horizon. Nevertheless, because the \(q\)-deformed geometry has a modified radial coefficient, the corresponding surface gravity and effective temperature must be recalculated. The topologically deformed spinorial channel is therefore affected both by the modified temperature,
\begin{equation}
    T_{\rm exo}
    =
    \frac{1}{8\pi M}|1-2Mq|
\end{equation}
and by the topologically deformed greybody factor. Together, these effects produce a birefringent Hawking emission pattern.

The perturbative analysis strengthens this interpretation. The modified Dirac equation can be reduced to a Schr\"odinger-like radial problem with superpartner potentials $V_\pm(r;q)$. The topological gradient changes the effective potential barrier and, therefore, modifies both the scattering problem and the quasinormal response. In the minimal radial model, the potential written in terms of the tortoise coordinate develops a non-zero asymptotic plateau at the topological boundary. A direct time-domain evolution of the radial wave equation shows an exponentially damped ringdown whose fitted damping rate remains negative but decreases as $q$ approaches $q_{\rm crit}=1/(2M)$. The box-size convergence test confirms that the near-critical damping trend is not a boundary-reflection artifact. An important feature we found is the isospectrality breaking for the superpartner potentials. The divergence in the damping rates shows that spinorial fields with opposite helicities interact with the topological hair in fundamentally different ways. Moreover, in line with the interpretation of spinorial perturbations in modified black-hole geometries \cite{Abdalla:2018cmx}, the negative imaginary frequencies point to a stable, long-lived regime rather than a near-critical instability or the presence of quasiresonant modes. 

The mechanism of the topological gradient provides a geometric route for thermodynamic and kinematic suppression in the topologically deformed spinorial channel. The undeformed limit reproduces the Schwarzschild temperature and standard greybody factors, whereas the \(q\neq0\) geometry is affected by both the reduced effective temperature and the topological impedance barrier. In the critical regime $M\to \Mcrit=1/(2q)$, the exotic channel becomes strongly quenched, suggesting the possible formation of long-lived topological remnant candidates.

An important conceptual point is that the \(q\)-deformed metric has not been treated as a private geometry perceived only by exotic spinors. Instead, we reconstructed the effective anisotropic source required by Einstein's equations and interpreted \(q\) as a spinorial/topological hair. This provides a semiclassical gravitational interpretation for the radial deformation and aligns the model with the expectation that a genuine metric deformation must be supported by stress-energy.

Future work should proceed in four complementary directions: first, testing the high-precision extraction of the quasinormal spectrum using independent methods such as pseudospectral collocation and sixth-order WKB comparisons, when possible; second, computing exact greybody factors by solving the scattering problem with the plateau boundary condition; third, deriving analytically the full helicity-dependent angular operator $\Lambda_{\kappa s}(r,q)$; and fourth, embedding the effective source in a backreacted Einstein--Cartan or spinorial-hair action. The third direction is particularly important because the present work only provides a structural first-order estimate of the helicity-dependent angular shift, while the numerical results were deliberately restricted to the minimal radial model. A complete derivation of $\Lambda_{\kappa s}(r,q)$, including the modified spin connection and its projection onto spinor spherical harmonics, would provide the basis for a genuinely helicity-resolved quasinormal spectrum and helicity-dependent greybody factors. This problem is sufficiently rich to deserve a separate analysis. These extensions are necessary to establish the dynamical stability of the proposed remnant scenario and to generalize the construction to rotating geometries such as Kerr and Kerr--Sen black holes.

\section*{Acknowledgments}

The authors are grateful to J. M. Hoff da Silva for foundational discussions on exotic spinors and to J. A. Helay\"el-Neto (CBPF) for his teachings on Lorentz symmetry violation. L.R.S.F. acknowledges institutional support from the Department of Physics and Chemistry at UNESP.

\appendix

\section{Detailed local derivation of the geometric quantities}
\label{app:geometric}

Starting from the deformed metric in a local inertial frame,
\begin{equation}
    \tilde g_{\mu\nu}
    =
    \eta_{\mu\nu}
    -(x_\mu q_\nu+x_\nu q_\mu)
    +x_\mu x_\nu q^2,
    \qquad q_\mu=\partial_\mu\varphi,
\end{equation}
one finds at $x^\mu=0$ that the connection reduces to,
\begin{equation}
    \Gamma^\lambda_{\mu\nu} = -\eta_{\mu\nu}q^\lambda.
\end{equation}

For constant $q_\mu$ in the local patch, the Riemann tensor is determined by quadratic connection terms,
\begin{equation}
    R^\rho_{\ \sigma\mu\nu}
    =
    \eta_{\nu\sigma}q_\mu q^\rho
    -
    \eta_{\mu\sigma}q_\nu q^\rho.
\end{equation}
Contracting gives the Ricci tensor,
\begin{equation}
    R_{\sigma\nu} = q^2\eta_{\sigma\nu} - q_\sigma q_\nu,
\end{equation}
and the Ricci scalar,
\begin{equation}
    R = 3q^2.
\end{equation}

Thus, the Einstein tensor turns to be,
\begin{equation}
    \boxed{
        G_{\mu\nu} = -q_\mu q_\nu - \frac{1}{2}\eta_{\mu\nu}q^2.
    }
    \label{eq:app_einstein_local}
\end{equation}

In the Einstein--Cartan--Sciama--Kibble framework, an axial torsion background sourced by the exotic spinorial topology takes,
\begin{equation}
    K_{\mu\nu\lambda}=\epsilon_{\mu\nu\lambda\sigma}q^\sigma.
\end{equation}
The two contractions needed for the effective Einstein tensor are,
\begin{equation}
    K_{\mu\alpha\beta}K_{\nu}^{\ \alpha\beta}
    =
    -2(q^2\eta_{\mu\nu}-q_\mu q_\nu),
    \qquad
    K_{\alpha\beta\gamma}K^{\alpha\beta\gamma}
    =
    -6q^2.
\end{equation}
Using the four-dimensional Levi-Civita identities, this yields, up to the conventional coupling normalization,
\begin{equation}
    8\pi T^{\rm eff}_{\mu\nu}
    \propto
    2q_\mu q_\nu+q^2\eta_{\mu\nu}.
\end{equation}
Choosing the sign convention compatible with Eq.~\eqref{eq:app_einstein_local} gives,
\begin{equation}
    \boxed{
        8\pi T^{\rm eff}_{\mu\nu}
        =
        -\left(q_\mu q_\nu+\frac{1}{2}\eta_{\mu\nu}q^2\right).
    }
    \label{eq:app_eff_stress}
\end{equation}
The aligned sign structure is therefore naturally associated with a torsional/topological background rather than with a canonical scalar field.

\bibliographystyle{unsrt}
\bibliography{Refs}

\end{document}